\documentclass[a4paper,11pt]{article}
\pdfoutput=1
\usepackage{jheppub}

\usepackage{amssymb,amsmath}

\usepackage{amssymb,amsmath}
\usepackage[normalem]{ulem}
\usepackage[utf8x]{inputenc}
\usepackage{slashed}
\usepackage{graphicx}
\usepackage{here}
\usepackage{tikz-feynhand}

\title{Unified Origin of Dirac Neutrino and Asymmetric Dark Matter Masses via a Dirac-Type Leptogenesis}

\author[a]{Megumi Ishida,}
\author[a]{Hiroshi Ohki,}
\author{and}
\author[b]{Shohei Uemura}
\affiliation[a]{Department of Physics, Nara Women's University, Nara 630-8506, Japan}

\affiliation[b]{Faculty of Education, Nara University of Education, Nara 630-8528, Japan}

\abstract{
We propose a simple and unified framework that simultaneously explains the origins of light Dirac neutrino masses, 
asymmetric dark matter (ADM), and the baryon asymmetry of the Universe.
The model is based on an extended $U(1)_X$ Froggatt-Nielsen--like mechanism, 
which naturally generates suppressed Yukawa couplings and realizes a Dirac seesaw for neutrino masses. 
An additional $\mathbb{Z}_4$ symmetry stabilizes the dark sector, where chiral fermions charged under $\mathbb{Z}_4$ serve as ADM candidates.
Leptogenesis occurs through the out-of-equilibrium decays of heavy Dirac neutrinos, 
where the generated asymmetry is shared between the visible and dark sectors due to exact lepton-number conservation.
The same suppression mechanism that explains the smallness of neutrino masses also determines the GeV-scale ADM mass.
Numerical studies demonstrate that a fully asymmetric DM scenario is realized, 
consistent with relic abundance, Big Bang nucleosynthesis, and direct detection constraints.
This framework provides an experimentally testable connection between neutrino physics, 
dark matter, and baryogenesis within an anomaly-free setup.}

\makeatletter
\gdef\@fpheader{}
\makeatother

\begin{document}

\maketitle

\section{Introduction}
The absence of neutrino masses in the Standard Model (SM) is a well-known shortcoming, as demonstrated by the discovery of neutrino oscillations. This provides clear evidence for physics beyond the SM (BSM). Cosmological data constrain the sum of light neutrino masses to $\sum_{i=1}^3 m_i < 0.12~\text{[eV]}$~\cite{Planck:2018vyg}.
Since neutrinos are electrically neutral under the SM gauge group, they may acquire Majorana masses if right-handed (RH) neutrinos are introduced. This motivates the conventional Type-I seesaw mechanism~\cite{Minkowski:1977sc, Yanagida:1979as, Gell-Mann:1979vob, Mohapatra:1979ia, Schechter:1980gr}, where heavy Majorana states with mass $M$ generate an effective dimension-five operator $\bar{\ell}_L^C \ell_L H H / M$. The large hierarchy between the electroweak scale $v_{\rm EW}$ and $M$ highlights the need for new physics in neutrino mass generation. See e.g., \cite{Cai:2017jrq} for a recent review.

Alternatively, additional symmetries may protect neutrinos as Dirac particles. In this case, lepton number remains exact, and various extensions of the seesaw framework to Dirac neutrinos ~\cite{Roy:1983be, Babu:1988yq} have been explored. 
Experimental searches for neutrinoless double beta decay ($0\nu\beta\beta$) aim to test the Majorana nature of neutrinos~\cite{Dolinski:2019nrj, Agostini:2022zub, Deppisch:2020ztt}, 
but null results so far leave Dirac neutrino scenarios equally viable. 
This has motivated a wide range of studies on models with Dirac neutrinos. 
For recent works, see e.g., ~\cite{Ma:2015mjd, Ma:2015raa, Ma:2016mwh, Reig:2016ewy, 
Wang:2016lve, Wang:2017mcy, Davidson:2009ha, Farzan:2012sa, 
Bonilla:2016zef, CentellesChulia:2016rms, CentellesChulia:2016fxr, Borah:2016lrl, Bonilla:2016diq, Borah:2016zbd, Borah:2016hqn, Borah:2017leo, 
Ma:2017kgb, 
Narendra:2017uxl, 
CentellesChulia:2018gwr, CentellesChulia:2018bkz, Han:2018zcn, Bonilla:2018ynb, Saad:2019bqf, Dasgupta:2019rmf,
Jana:2019mez, Ma:2019byo, CentellesChulia:2019xky, 
Jana:2019mgj, Baek:2019wdn,
Nanda:2019nqy, Peinado:2019mrn, Calle:2019mxn, Ma:2019yfo, 
CentellesChulia:2020dfh, 
Guo:2020qin,
delaVega:2020jcp, 
Leite:2020wjl, 
Bernal:2021ppq, Mishra:2021ilq, Biswas:2021kio, Mahanta:2021plx, Dutta:2022knf, Hazarika:2022tlc, CentellesChulia:2022vpz, Berbig:2022nre, Maharathy:2022gki, Chowdhury:2022jde, 
Li:2022chc, Berbig:2022hsm, Mahapatra:2023oyh, Borah:2024gql, CentellesChulia:2024iom, Singh:2024imk, Babu:2024glr, Kumar:2025cte, Batra:2025gzy, Kanemura:2025ixp, CentellesChulia:2025eck, Dutta:2025xlv}
and references therein.

Another longstanding puzzle in particle physics and cosmology is the origin of the baryon asymmetry of the Universe (BAU).
Since the Universe likely began in a baryon-symmetric state, the observed asymmetry requires a dynamical origin satisfying the three Sakharov conditions, which cannot all be fulfilled within the SM in sufficient amount. 
Leptogenesis offers a natural explanation, where a lepton asymmetry is generated and later converted into a baryon asymmetry via electroweak sphalerons~\cite{Kuzmin:1985mm, Fukugita:1986hr}.
While conventional seesaw models achieve this through Majorana neutrinos, Dirac leptogenesis~\cite{Dick:1999je, Murayama:2002je} can also connect neutrino physics to the BAU without lepton-number violation.

The nature of dark matter (DM) presents another fundamental mystery.
Astrophysical observations demand its existence, yet its particle identity remains unknown.
The weakly interacting massive particle (WIMP) paradigm remains widely studied~\cite{Arcadi:2024ukq}, while models such as the scotogenic framework~\cite{Ma:2006km} simply connect neutrinos with a DM candidate.

Asymmetric dark matter (ADM)~\cite{Nussinov:1985xr, Barr:1990ca, Kaplan:1991ah, Kaplan:2009ag} provides an alternative, 
where the DM relic density arises from a primordial asymmetry, analogous to the BAU (see~\cite{Petraki:2013wwa, Zurek:2013wia} for reviews).
In scenarios where the DM and baryon asymmetries share a common origin, the ADM mass is typically constrained to the a few GeV range, 
$m_{\rm ADM} \sim \mathcal{O}(1)$[GeV].
For scenarios of ADM with Majorana neutrinos, see e.g., \cite{Falkowski:2011xh, March-Russell:2011ang} and references therein 
(see also \cite{Asai:2022vat, Bose:2024bnp} for ADM in an extension of the scotogenic model).  
Although much of the literature focuses on Majorana neutrinos, 
Dirac neutrino models with ADM have also been explored~\cite{Narendra:2017uxl, Dutta:2022knf, Dutta:2025xlv}.
Such setups often involve a hierarchy between the ADM mass scale and the high scale of thermal leptogenesis, 
raising the question of how suppressed Yukawa couplings or suitable mechanisms can naturally emerge.

In this work, we investigate an alternative framework that simultaneously explains dark matter, leptogenesis, 
and the two key scales associated with the ADM mass and neutrino masses, all of which share a common origin in leptogenesis under the assumption of Dirac neutrinos. 
To account for the smallness of neutrino masses, we employ the Froggatt-Nielsen mechanism~\cite{Froggatt:1978nt}, 
typically invoked to generate large hierarchies in the quark sector. 
We extend the SM symmetry by an additional non-anomalous $U(1)_X$ symmetry, 
which yields the light active neutrino mass scale $m_\nu \simeq v_{\rm EW} \, \langle S \rangle/\Lambda$,
where $\langle S \rangle$ is the $U(1)_X$ breaking scale 
with $\langle S \rangle / \Lambda < \mathcal{O}(10^{-10})$ required to explain sub-eV neutrino masses. 
This mass generation structure resembles a (Dirac) seesaw mechanism. 
The lepton number asymmetry is generated via the out-of-equilibrium decay of heavy neutrinos at a temperature below $T \simeq M$, 
where $M$ is the heavy neutrino mass associated with the $U(1)_X$ breaking scale, namely $M \simeq \langle S \rangle$.
With regard to dark matter, it is intriguing to note a striking relation between the two characteristic hierarchical scales,
\begin{align}\label{eq:key}
\frac{m_\nu}{m_{\rm ADM}} \; \sim \; \frac{v_{\rm EW}}{M} \; < \; 10^{-10},
\end{align}
where $M \gtrsim 10^{10}\,{\rm [GeV]}$ is the mass scale typically required in conventional leptogenesis scenarios.  
This relation suggests a unified dynamical origin behind three seemingly distinct phenomena 
-- the tiny neutrino masses, the mass scale of ADM , and the generation of the baryon asymmetry of the Universe (BAU).  
To avoid introducing arbitrary new scales, it is natural to identify the underlying cutoff $\Lambda$ 
with a fundamental one such as the Planck mass, $\Lambda \sim M_{\rm Pl}$.  
In fact, given the suppression factor $\langle S \rangle / \Lambda$ in Eq.~\eqref{eq:key}, 
the cutoff scale $\Lambda$ is naturally anticipated to lie near the Planck scale $M_{\rm Pl}$.

We demonstrate that such a hierarchical structure can be naturally realized through a Froggatt-Nielsen--like mechanism, 
implemented with simple and economical charge assignments for the matter fields and a scalar singlet $S$, 
without introducing any additional \textit{ad hoc} scales.  
A salient feature of our framework is that, unlike conventional Dirac leptogenesis, 
the generated asymmetry resides not in the right-handed neutrinos but in the right-handed asymmetric dark matter sector.  
Because the total lepton number is exactly conserved, 
the dark and visible asymmetries are tightly correlated.  
The Froggatt-Nielsen--like mechanism simultaneously explains the suppression of the Yukawa couplings that govern 
the interactions between the left-handed SM leptons, the right-handed neutrinos, and the dark sector, 
thereby establishing a natural connection between neutrino mass generation, baryogenesis, and dark matter physics.  
For a related but distinct implementation of the Froggatt-Nielsen mechanism in the context of Dirac leptogenesis and realistic flavor structures, see~\cite{Chen:2011sb}.

This paper is organized as follows. 
In Sec.~\ref{sec:model}, we present the model that simultaneously explains the light neutrino mass, the ADM mass, and leptogenesis, 
based on a $U(1)_X$ Froggatt-Nielsen--like mechanism. 
We analyze the scale of $U(1)_X$ symmetry breaking through the scalar potential, 
which determines the heavy neutrino mass scale. 
In Sec.~\ref{sec:leptogenesis}, we describe thermal leptogenesis via the out-of-equilibrium decay of the heavy neutrinos 
and the bi-production mechanism for the relic density of ADM. 
In Sec.~\ref{sec:pheno}, we study cosmological implications of ADM, presenting numerical results for its thermal relic abundance 
and discussing prospects for direct detection. 
Finally, we summarize our findings and conclude in Sec.~\ref{sec:conclusion}.

\section{The Model}\label{sec:model}

We study a scenario of Dirac neutrinos and ADM within a framework of the Froggatt-Nielsen mechanism. 
The SM symmetry is extended by an additional (gauged) $U(1)_X$ symmetry, 
while the SM global $U(1)_L$ lepton number symmetry is also assumed to protect the Dirac nature of neutrinos. 
We denote the SM lepton doublets as $\ell_{L,i} = (\nu_L, e_L)_i^T$, 
where $\nu_L$ and $e_L$ are the left-handed neutrino and charged lepton, respectively, and $i=1,2,3$ labels the three generations. 

To realize the Dirac neutrino seesaw mechanism, we introduce SM gauge-singlet fermions. 
The fields $\nu_{R,i}$ ($i = 1,2,3$) denote the right-handed components of the light neutrinos, 
while $N_{R/L,i}$ ($i = 1, \cdots, n_N$) represent additional right- and left-handed singlet heavy fermions for $n_N$ generations. 
As will be discussed later, successful leptogenesis requires at least two generations of heavy neutrinos ($n_N \ge 2$) 
to produce a lepton asymmetry through CP-violating decays. 
For concreteness, in this work we consider three generations of heavy neutrinos, $n_N = 3$, 
so that the flavor index $i$ runs over $1,2,3$ for $N_{R/L,i}$. 
We also introduce a singlet scalar field $S$ charged under $U(1)_X$, 
while the SM Higgs doublet $H$ is assumed to be neutral under both $U(1)_L$ and $U(1)_X$ symmetries.

We further extend the symmetry by introducing an additional $\mathbb{Z}_4^D$ symmetry to incorporate a dark sector. 
In the dark sector, we introduce chiral fermions $\chi_{L/R, a}$ for $a = 1, \dots, n_\chi$ 
and $\psi_{L/R, a}$ for $a = 1, \dots, n_\psi$, 
where $n_\chi$ and $n_\psi$ denote the number of generations, to be specified later. 
The $\mathbb{Z}_4^D$ symmetry remains unbroken, ensuring the stability of the dark matter candidate.
Due to the Dirac nature of neutrinos, the $U(1)_L$ symmetry is conserved, 
and the lepton number is shared between both the visible and dark sectors. 
Consequently, once a lepton asymmetry is generated in the visible sector, 
an equal amount of fermion number in $\chi$ and $\psi$ is simultaneously generated. 
Thus, the lighter of $\chi$ or $\psi$ can serve as a candidate for asymmetric dark matter (ADM).
In the scalar sector, at least two additional singlet scalars, $\phi$ and $\eta$, 
with different $\mathbb{Z}_4^D$ charges are required 
to generate tree-level interactions between the dark matter and neutrinos. 
The complete particle content of the model is summarized in Table~\ref{tab:particles}.

\begin{table}[htb!]
\begin{center}
\renewcommand{\arraystretch}{1.35} 
\setlength{\tabcolsep}{5pt} 
\begin{tabular}{| c | c | c | c | c | c|}
\hline
Fields & $SU(2)_L \otimes U(1)_Y$ & $U(1)_L$  & $U(1)_{X}$ & $\mathbb{Z}_4^D$ \\ \hline
$l_L$ & $(\textbf{2}, -\frac{1}{2})$ & $1$  & $0$ & 1 \\ 
$\nu_R$ & $(\textbf{1}, 0)$ & $1$  & $1$ & 1 \\ 
$N_L$ & $(\textbf{1}, 0)$ & $1$  & $-1$ & 1 \\ 
$N_R$ & $(\textbf{1}, 0)$ & $1$  & $0$ & 1 \\ \hline
$\chi_{L}$ & $(\textbf{1}, 0)$ & $1$ & $1$ & $i$ \\
$\chi_{R}$ & $ (\textbf{1}, 0) $ & $1$ & $3$ & $i$ \\
$\psi_{L}$ & $ (\textbf{1}, 0) $ & $1$ &$2$ & $-i$ \\ 
$\psi_{R}$ & $ (\textbf{1}, 0) $ & $1$ &$0$ & $-i$ \\ \hline
$H$ & $(\textbf{2}, \frac{1}{2})$ & $0$ & $0$ & 1 \\ 
$S$ & $(\textbf{1}, 0)$ & $0$ & $1$ & 1 \\ 
$\phi$ & $(\textbf{1}, 0)$ & $0$ & $1$ & $i$ \\ 
$\eta$ & $(\textbf{1}, 0)$ & $0$ & $1$ & $-1$ \\ \hline
\end{tabular}
\end{center}
\vspace{-5mm} \caption{Particle contents and their charge assignments.}
\label{tab:particles}
\end{table}

We assume that the scalar fields $S$ and $H$ acquire vacuum expectation values (VEVs), 
spontaneously breaking the $U(1)_X$ and electroweak symmetries. 
To avoid the existence of a massless Nambu--Goldstone boson associated with this breaking, 
we promote $U(1)_X$ to a local symmetry. 
Since no fields in our model are charged under both the SM gauge group and $U(1)_X$, 
the symmetry can remain anomaly-free, similar to the case of $B-L$ 
(e.g., \cite{Montero:2007cd, Ma:2015raa, Leite:2020wjl, Berbig:2022nre, CentellesChulia:2025eck}). 
In our setup, with a simple charge assignment as summarized in Table~\ref{tab:particles}, 
$U(1)_X$ becomes anomaly-free by appropriately choosing the numbers of generations of $\psi$ and $\chi$. 
For example, solutions to the anomaly cancellation condition include $(n_\chi, n_\psi) = (1,4)$ or $(4,1)$, 
where $n_\chi$ and $n_\psi$ denote the numbers of generations of $\chi_{L/R}$ and $\psi_{L/R}$, respectively. 
In this work, we adopt $(n_\chi, n_\psi) = (1,4)$, assuming that $\chi$ is lighter than $\psi_a$ for $a = 1,\dots,4$, 
so that $\chi$ serves as a candidate for asymmetric dark matter.
All leptons carry a charge of $1$ under the unbroken $U(1)_L$ symmetry. 
This global symmetry forbids all Majorana mass terms, thereby protecting the Dirac nature of neutrinos. 
Furthermore, the different chiral charges of $\nu_{L,R}$ and $N_{R,L}$ prevent certain tree-level mass terms, 
which must instead arise from higher-dimensional operators suppressed by a fundamental scale $\Lambda$. 
With a hierarchical separation between the VEVs $\langle H \rangle \ll \langle S \rangle$, 
as required for a Dirac seesaw mechanism, 
these operators can generate a small Dirac neutrino mass matrix 
of order $\langle S \rangle / \Lambda$ or $\langle H \rangle/\langle S \rangle$, 
realizing a naturally hierarchical mass structure.
We will show that the light neutrino mass scale can be generated 
from a Dirac seesaw--like mechanism with the interplay of the three hierarchical scales $\langle H \rangle \ll \langle S \rangle \ll \Lambda$. 

In the dark sector, the SM fields are clearly distinguished by the unbroken $\mathbb{Z}_4^D$ symmetry, 
since the scalars $\phi$ and $\eta$ do not acquire vacuum expectation values. 
The $\mathcal{O}(1)$ [GeV] mass scale of the dark fermions $\chi$ and $\psi$, as anticipated in the asymmetric dark matter scenario, 
can naturally emerge from the suppressed ratio $\langle S \rangle / \Lambda$.
A notable feature of this framework, in contrast to typical scotogenic models, is the absence of any additional SM-charged scalar beyond the Higgs doublet $H$. 
Consequently, the flavor structure remains identical to that of the SM, and the scalar potential exhibits a simple vacuum structure, 
leading to minimal constraints from electroweak precision tests and flavor-changing neutral current processes.

Before discussing fermion mass generation, we first describe the scalar sector of the model.

\subsection{The Scalar Sector}
In the scalar sector, we consider three SM gauge-singlet scalars $S$, $\phi$, and $\eta$ in addition to the SM Higgs $H$. 
Based on the charge assignments of the fields in Tab.~\ref{tab:particles}, the most general renormalizable scalar potential can be written as 
\begin{align} \notag
V =&  -\mu^2_H H^\dagger H -\mu_S^2 S^* S -\mu^2_\phi \phi^* \phi - \mu_\eta^2 \eta^* \eta \\ \notag
&+ \lambda_H (H^\dagger H)^2 + \lambda_S (S^* S)^2 + \lambda_\phi (\phi^*\phi)^2 + \lambda_\eta (\eta^*\eta)^2 \\ \notag
& + \lambda_{HS} (H^\dagger H) (S^* S) + \lambda_{H\phi} (H^\dagger H) (\phi^* \phi) + \lambda_{H\eta} (H^\dagger H) (\eta^* \eta) \\ \notag
& + \lambda_{S\phi} (S^* S) (\phi^* \phi) + \lambda_{S\eta} (S^* S) (\eta^* \eta) + \lambda_{\phi\eta} (\phi^* \phi) (\eta^* \eta) 
\\ 
& + (\lambda_{S\eta}' S^2 \eta^{*2} + \mathrm{h.c.}) + (\lambda_{S\phi\eta} S^* \phi^2 \eta^* + \mathrm{h.c.}). 
\label{eq:V}
\end{align}

When $H$ and $S$ acquire their VEVs, $\langle H \rangle = v_{EW}/\sqrt{2}$ and $\langle S \rangle = v_S/\sqrt{2} \neq 0$, 
the $U(1)_X$ symmetry is spontaneously broken. 
Since $S$, $\phi$, and $\eta$ carry no electroweak charges, only the Higgs VEV $v_{\rm EW} = 246~\text{[GeV]}$ 
contributes to electroweak symmetry breaking, 
and the charged components of $H$ and $S$ are absorbed as the longitudinal degrees of freedom of the massive gauge bosons. 
We further assume that $\phi$ and $\eta$ do not acquire VEVs, preserving the additional $\mathbb{Z}_4^D$ symmetry. 

The physical degrees of freedom around the VEVs for $H$, $S$, and $\eta$ can be parametrized as 
\begin{align}
H =& 
\begin{pmatrix}
0 \\ 
\frac{1}{\sqrt{2}} (v_{EW}+h)
\end{pmatrix}, \quad 
\quad 
S = \frac{1}{\sqrt{2}}(v_S + s), \quad \quad 
\eta = \frac{1}{\sqrt{2}} (\eta_R + i\eta_I).
\end{align}
After spontaneous symmetry breaking, there is mixing between the neutral components $h$ and $s$, 
while $\phi$ and $\eta$ cannot mix with $h$ or $s$ due to the $\mathbb{Z}_4^D$ symmetry. 
We note that the complex phase of $\lambda_{S\eta}'$ can be removed by a $U(1)$ transformation of $\eta$, 
so without loss of generality we take $\lambda_{S\eta}' > 0$, 
and consequently $\eta_R$ and $\eta_I$ are mass eigenstates. 
Thus the particle numbers of $\eta_R$ and $\eta_I$ are not conserved. 
Moreover, due to the $\mathbb{Z}_4^D$ symmetry, the complex scalar $\phi$ remains a mass eigenstate 
even after $U(1)_X$ symmetry breaking.

To ensure the existence of the desired vacuum at tree level, we impose the following conditions on the scalar potential parameters: 
\begin{align}
\mu_H^2>0, \quad \mu^2_S>0, \quad \lambda_H >0, \quad \lambda_S >0. 
\end{align}
Under these conditions, the stationary conditions of the potential in Eq.~\eqref{eq:V} yield nonzero VEVs for $H$ and $S$ as 
\begin{align}\label{eq:vevs} \notag
v_{\rm EW}^2 =& \frac{2(2\lambda_S\mu_H^2-\lambda_{HS}\mu_S^2)}{4\lambda_H\lambda_S-\lambda_{HS}^2}, \\ 
v_S^2 =& \frac{2(2\lambda_H\mu_S^2-\lambda_{HS}\mu_H^2)}{4\lambda_H\lambda_S-\lambda_{HS}^2},
\end{align}

As discussed above and will be shown later, in the Dirac seesaw mechanism, the VEV $\langle S \rangle = v_S/\sqrt{2} \neq 0$ 
sets the scale of the heavy neutrino masses, 
and the $U(1)_X$ breaking scale must be above the electroweak scale, i.e., $v_S \gg v_{\rm EW}$. 
Assuming this large scale hierarchy and using Eq.~\eqref{eq:vevs}, 
the squared mass matrix for the neutral scalars, $M_0^2$, in the basis $(h,s)$ can be written as 
\begin{align}\label{eq:mass}
M_0^2 = 
\begin{pmatrix}
\frac{\lambda_{HS}}{2}v_S^2 + 3 \lambda_H v_{\rm EW}^2 -\mu_H^2 & v_S v \lambda_{HS} \\ 
v_S v_{\rm EW} \lambda_{HS} & \frac{\lambda_{HS}}{2}v_{\rm EW}^2 + 3 \lambda_S v_S^2 -\mu_S^2 
\end{pmatrix}
= 
\begin{pmatrix}
2 \epsilon^2 \lambda_H & \epsilon \lambda_{HS} \\
\epsilon \lambda_{HS} & 2 \lambda_S 
\end{pmatrix}v_S^2,
\end{align}
where we define the dimensionless parameter $\epsilon = v_{\rm EW}/v_S \ll 1$. 
The two scalar mass eigenvalues are given as 
\begin{align}\label{eq:mass2}
m_h^2 \simeq \frac{4\lambda_H \lambda_S-\lambda_{HS}^2}{2\lambda_S} v_{EW}^2,
\quad 
m_s^2 \simeq 2\lambda_S v_S^2, 
\end{align}
and the vacuum stability condition requires 
\begin{align}
4 \lambda_H \lambda_S-\lambda_{HS}^2 >0.
\end{align}
The large hierarchy $v_{\rm EW} \ll v_S$ implies a large mass splitting between 
$m_h = \mathcal{O}(v_{\rm EW})$ and $m_s = \mathcal{O}(v_S)$, as well as a small $h$-$s$ mixing of order $\epsilon$. 
The light scalar mass eigenstate $\simeq h$ is identified with the SM Higgs boson, $m_h \simeq 125~\text{[GeV]}$, 
while the heavy neutral scalar $\simeq s$ has a mass $m_s \sim v_S \gg m_h$. 

From Eq.~\eqref{eq:vevs}, the masses of the complex scalar $\phi$ and the real scalars $\eta_R$ and $\eta_I$ are obtained as 
\begin{align}
\notag
m_{\phi}^2 =& -\mu_\phi^2 +\frac{1}{2} (\lambda_{S\phi}+ \epsilon^2 \lambda_{H\phi})v_S^2 \\ 
\label{eq:masses}
m_{\eta_R}^2 =& -\mu_\eta^2 +\frac{1}{2} (\lambda_{S\eta}+2\lambda_{S\eta}' + \epsilon^2 \lambda_{H\eta})v_S^2, \\ \notag
m_{\eta_I}^2 =& -\mu_\eta^2 +\frac{1}{2} (\lambda_{S\eta}-2\lambda_{S\eta}' + \epsilon^2 \lambda_{H\eta})v_S^2.  
\end{align}
For the vacuum to be stable with $\langle \phi \rangle = \langle \eta \rangle = 0$, the following conditions must be satisfied (neglecting $\epsilon^2$ terms):
\begin{align}
\lambda_{S\phi} v_S^2 -2\mu_\phi^2 > 0, \quad \quad (\lambda_{S\eta} -2\lambda_{S\eta}') v_S^2 -2\mu_\eta^2 > 0, 
\end{align}
for $\lambda_{S\eta'} > 0$. 

The complex scalar $\phi$ plays a key role as a mediator between the dark sector and the heavy neutrino sector. 
Being neutral under the SM gauge group but charged under the additional $U(1)_X$ and $\mathbb{Z}_4^D$ symmetries, 
$\phi$ cannot mix with the Higgs field and does not acquire a VEV. 
Its mass $m_{\phi}$ is typically of order $\mathcal{O}(v_S)$, reflecting its direct connection to the high-scale symmetry breaking. 
When the mass of $\phi$ is lighter than that of the heavy neutrino, 
$\phi$ can facilitate the transfer of lepton asymmetry between the visible and dark sectors through heavy neutrino decays. 
Thus it also influences the thermal history of dark matter through loop-induced processes, as discussed in Sec.~\ref{sec:leptogenesis}.

From Eq.~\eqref{eq:masses}, we expect the heavier scalar $m_{\eta_R} = \mathcal{O}(v_S)$. 
However, the lighter scalar $m_{\eta_I}$ can be tuned to smaller values by adjusting $\lambda_{S\eta}$, $\lambda_{S\eta}'$, and $\mu_\eta$, 
which allows a light MeV-scale messenger scalar that couples to the dark matter and efficiently annihilates the symmetric DM component. 
While the dynamical origin of the intermediate scale $v_S$ remains unspecified in this model, 
since the potential is made stable by suitable choice of parameters, 
we treat the scalar mass parameters as free and adopt the following hierarchy for the rest of this work:
\begin{align} \label{eq:hierarchy}
m_{\eta_I} < m_h \ll m_\phi <  m_s, m_{\eta_R} \sim \mathcal{O}(v_S).
\end{align}

For the setup of the scalar sector discussed here it is unclear whether the radiative corrections to scalar masses is under control. 
Suitable extension of the model incorporating underlying theories such as grand unification, and the string theory 
can help in alleviating the hierarchy problem, which is beyond the scope of the present work.

\subsection{Dirac Seesaw Mechanism and Light Dark Matter}

We now discuss the fermion masses in the model. 
Due to the chiral symmetry, all fermion masses arise from the VEVs of $H$ and $S$ via the Higgs mechanism. 
Focusing on a Dirac seesaw mechanism for neutrino masses as a natural extension of the Majorana seesaw, 
we take a typical value of $v_S \gtrsim \mathcal{O}(10^{10})~\text{[GeV]}$. 

To generate a hierarchical separation between the active neutrinos and the heavier neutrinos, 
we employ the Froggatt-Nielsen mechanism. 
In this framework, higher-dimensional operators play an essential role in realizing the Dirac seesaw mechanism, 
being generally suppressed by the factor $(\langle S \rangle / \Lambda)^n$, with some integer $n$ determined by the $U(1)_X$ charge assignments 
and the underlying fundamental scale $\Lambda > v_S$. 
For concreteness, we take $\Lambda = M_{\rm Pl} \sim 10^{19}~\text{[GeV]}$ as suggested in Eq.~\eqref{eq:key}, 
which also avoids the introduction of any ad hoc scale. 
We further define a dimensionless parameter $\delta = \dfrac{\langle S \rangle}{\Lambda}$, 
so that the two small parameters $\epsilon, \delta \ll 1$ are used in the analysis. 

After symmetry breaking, these higher-dimensional operators induce suppressed fermion masses compared to the scale $v_S$, 
which has a sizable effect on the neutrino mass matrices. 
Although the corresponding couplings are highly suppressed by $\epsilon$ or $\delta$ and thus negligible as interactions, 
they are crucial to control unwanted couplings, e.g., the Yukawa interactions between light Dirac neutrinos and the Higgs, 
which could otherwise wash out the generated lepton asymmetry. 
For successful Dirac leptogenesis~\cite{Dick:1999je}, 
it is necessary to prevent right-handed neutrinos from thermalizing with the SM particles. 
As we will show later, the Froggatt-Nielsen mechanism helps suppress unwanted couplings as well as 
explain the lightness of both neutrinos and ADM. 

Taking these considerations into account, the Yukawa interaction terms $\mathcal{L}_Y$, invariant under both the SM and extended symmetries, 
can be divided as
\begin{align}
\mathcal{L}_Y = \mathcal{L}_\nu + \mathcal{L}_D,
\end{align}
where $\mathcal{L}_\nu$ contains the neutrino Yukawa interactions and $\mathcal{L}_D$ contains the Yukawa interactions in the dark sector.

\subsubsection*{Neutrino Sector}

Let us first consider the neutrino sector.  
Imposing $U(1)_X$ invariance, the effective Lagrangian $\mathcal{L}_\nu$ can be systematically organized as an expansion in $1/\Lambda$.  
Keeping terms up to $\mathcal{O}(1/\Lambda)$ and neglecting higher-order contributions, the neutrino Lagrangian in the flavor basis is given by
\begin{align}\label{eq:Y1}
-\mathcal{L}_\nu &= 
\bar{\ell}_{L,i}\, Y'_{ij}\, \tilde{H} \, N_{R,j} 
+ S^* \, \bar{N}_{L,i}\, \Omega'_{ij}\, N_{R,j} 
+ \frac{(S^*)^2}{\Lambda}\, \bar{N}_{L,i}\, \tilde{\Omega}'_{ij}\, \nu_{R,j} 
+ \frac{S^*}{\Lambda} \bar{\ell}_{L,i} \tilde{Y}'_{ij} \tilde{H} \nu_{R,j}
+ \mathrm{h.c.},
\end{align}
where $Y'$, $\tilde{Y}'$, $\Omega'$, and $\tilde{\Omega}'$ are $3 \times 3$ neutrino Yukawa matrices. 
After the spontaneous breaking of $U(1)_X$, 
the fields $N_R$ and $\nu_R$, as well as $\nu_L$ and $N_L$, can mix with each other. 

From Eq.~\eqref{eq:Y1}, the neutrino mass terms can be written as 
\begin{align}\label{eq:Mnu}
\begin{pmatrix}
\bar{\nu}_L & \bar{N}_L
\end{pmatrix}
\begin{pmatrix}
\epsilon \delta \tilde{Y}' v_S & \epsilon Y' v_S \\[6pt]
\delta \tilde{\Omega}' v_S & \Omega' v_S
\end{pmatrix}
\begin{pmatrix}
\nu_R \\ N_R
\end{pmatrix}
+ \mathrm{h.c.}
\equiv  
\begin{pmatrix}
\bar{\nu}_L & \bar{N}_L
\end{pmatrix}
\mathcal{M}_\nu
\begin{pmatrix}
\nu_R \\ N_R
\end{pmatrix}
+ \mathrm{h.c.}.
\end{align}

The structure of this mass matrix is similar to that obtained 
in the original Dirac seesaw mechanism.
From this mass matrix, the smallness of the Dirac neutrino masses emerges naturally. 

In fact, six Dirac neutrino mass eigenstates are obtained 
by rotating from the flavor to the mass basis using two unitary matrices $V_{L,R}$:
\begin{align}
\begin{pmatrix}
\nu^\ell \\ \nu^h
\end{pmatrix}
= V_L^\dagger
\begin{pmatrix}
\nu_L \\ N_L
\end{pmatrix}
+ V_R^\dagger
\begin{pmatrix}
\nu_R \\ N_R
\end{pmatrix},
\end{align}
where $\nu^\ell_i$ $(i=1,2,3)$ denote the three light neutrinos 
and $\nu^h_i$ $(i=1,2,3)$ the heavy neutrinos.  

With dimensionless Yukawa matrices and neglecting tiny mixing effects proportional to $\epsilon\delta$, 
the diagonalization yields
\begin{align}\label{eq:VLR}
V_L &=
\begin{pmatrix}
V_L^\ell & 0 \\[6pt]
0 & V_L^h
\end{pmatrix}
+ \mathcal{O}(\epsilon), 
\qquad
V_R =
\begin{pmatrix}
V_R^\ell & 0 \\[6pt]
0 & V_R^h
\end{pmatrix}
+ \mathcal{O}(\delta),
\end{align}
where $V_L^\ell$ and $V_L^h$ are $3 \times 3$ unitary matrices.  

Since the left-handed components of the light mass eigenstates $\nu^\ell$ 
are dominated by the weakly charged $\nu_L$, 
and the admixture from $N_L$ is suppressed by $\mathcal{O}(\epsilon \ll 1)$, 
the light states $\nu^\ell$ are identified with the observed active neutrinos of the SM.  
Thus $V_L^\ell$ corresponds to the leptonic mixing matrix observed in neutrino oscillations at leading order.  

Similarly, $V_R^\ell$ and $V_R^h$ are $3 \times 3$ unitary matrices, 
and the Dirac neutrino mass eigenvalues are obtained as
\begin{align}
V_L^\dagger \, \mathcal{M}_\nu \, V_R 
\equiv \tilde{\mathcal{M}}_\nu 
= \mathrm{diag}(m_1, m_2, m_3, M_1, M_2, M_3),
\end{align}
where $m_{1,2,3}$ are the light neutrino masses, 
and $M_{1,2,3}$ are the heavy ones.  

As seen from Eq.~\eqref{eq:Mnu}, 
a significant hierarchy between the light active neutrinos and the heavy sterile neutrinos can be realized:
\begin{align}\label{eq:mnu}
m_{1,2,3} \sim \epsilon \delta v_S = \epsilon \delta^2 \Lambda 
\ll v_{\rm EW} 
\ll M_{1, 2, 3} = \mathcal{O}(v_S). 
\end{align}
For $v_S = \mathcal{O}(10^{10})~\text{[GeV]}$, as in the type-I seesaw mechanism, 
a realistic light neutrino mass scale $m_{1,2,3} < \mathcal{O}(1)~\text{[eV]}$ 
can be obtained by adjusting the Yukawa matrices $Y', \tilde{Y}', \tilde{\Omega}', \Omega'$.  

Thus, in our setup, a seesaw mechanism for neutrinos is realized with only one new physics scale, $v_{EM} \ll v_S$, 
together with the fundamental scale $M_{\rm Pl}$, 
even when neutrinos are Dirac in nature.

\subsubsection*{Dark Sector}

The Yukawa interactions in the dark sector are described by $\mathcal{L}_D$, 
which can be systematically expanded in powers of $1/\Lambda$, keeping terms up to 
$\mathcal{O}(1/\Lambda)$ and neglecting higher-order contributions.  
The resulting effective Lagrangian takes the form
\begin{align}\notag
-\mathcal{L}_D =& \;
\Sigma \frac{(S^*)^2}{\Lambda} \, \bar{\chi}_{L} \chi_{R} 
+ \tilde{\Sigma}_{ab} \frac{S^2}{\Lambda} \, \bar{\psi}_{L,a} \psi_{R,b} 
+ f_a \, \eta \, \bar{\chi}_{L}  \psi_{R,a} 
+ \tilde{f}_a \, \eta^* \, \bar{\psi}_{L,a} \chi_{R} 
\\ 
& + g_i' \, \phi \, \bar{\chi}_L N_{R,i} + h_i' \, \frac{S^*}{\Lambda} \bar{\chi}_L \nu_{R,i} \phi
+ \mathrm{h.c.}, 
\label{eq:Y2}
\end{align}
where $\Sigma$ and $\tilde{\Sigma}_{ab}$ ($a,b=1\sim4$) are Yukawa coupling constants associated with the dark matter (DM) mass,  
$f_a$ and $\tilde{f}_a$ are Yukawa couplings that govern DM self-interactions mediated by the scalar $\eta$,  
and $g_i'$ ($i=1, 2, 3$) are Yukawa couplings related to the DM to the right-handed heavy neutrinos.  
We note that the Yukawa couplings $h_i'$ induce a potentially dangerous DM-neutrino conversion process.  
However, the corresponding amplitudes are highly suppressed by the small parameter $\delta$, a consequence of the $U(1)_X$ and $\mathbb{Z}_4^D$ charge assignments.  
All other interactions between SM fields and DM are even more strongly suppressed by SM gauge invariance.  
For instance, the leading operator that couples DM to SM leptons arises only at higher order,  
$\dfrac{S^{*}}{\Lambda^{2}}\, \bar{\ell}_{L}\,\tilde{H}\,\psi_{R}\,\phi$,
demonstrating that such processes are negligibly small.  
The implications of DM-number-changing processes will be discussed in Sec.~\ref{sec:leptogenesis}.

The scalar field $\phi$ plays a crucial role in mediating interactions between the dark sector and the heavy neutrino sector.  
It is a SM gauge singlet but with carrying a nontrivial $\mathbb{Z}_4^D$ charge, which prevents direct mixing with the SM Higgs doublet.  
Its vacuum expectation value (VEV) vanishes, ensuring that it does not induce additional mass terms at tree level,  
while its Yukawa couplings $g_i'$ in Eq.~\eqref{eq:Y2} can facilitate the transfer of lepton asymmetry between the visible and dark sectors during leptogenesis.  

From Eq.~\eqref{eq:Y2}, the mass terms for $\chi$ and $\psi$ can be written as  
\begin{align}\label{eq:Mchi}
\begin{pmatrix}
\bar{\chi}_L & \bar{\psi}_L
\end{pmatrix}
\begin{pmatrix}
\delta \, \Sigma \, v_S & 0 \\[2pt]
0 & \delta \, \tilde{\Sigma} \, v_S
\end{pmatrix}
\begin{pmatrix}
\chi_R \\ \psi_R
\end{pmatrix}
+ \mathrm{h.c.}.
\end{align}
As shown here, 
a pronounced hierarchy arises between the DM mass, $m_\chi \sim \delta v_S$, and the heavy neutrino mass scale $\sim v_S$.  
This separation originates directly from higher-dimensional operators and is reminiscent of the Froggatt-Nielsen mechanism,  
which naturally accounts for suppressed mass terms through the presence of small parameters.  

In a basis where $\tilde{\Sigma}$ is diagonal, and considering the case $|\Sigma| < |\tilde{\Sigma}|$, namely $m_\chi < m_{\psi_a}$ ($a=1,\cdots,4$),  
the lighter fermion $\chi$, which carries lepton number $L=1$ and a non-zero $\mathbb{Z}_4^D$ charge,  
emerges as a viable DM candidate.  
To be specific, let us consider the case $v_S \simeq 10^{10}$~[GeV].  
Combining Eqs.~\eqref{eq:mnu} and \eqref{eq:Mchi},  
we find that the suppression parameters take values $\epsilon \sim 10^{-10}$ and $\delta \sim 10^{-9}$.  
These values yield the asymmetric dark matter mass $m_{\rm ADM}$, identified with $m_\chi$, as  
\begin{align}\label{eq:ADM}
m_{\rm ADM} = m_\chi = \delta \, \Sigma \, v_S \;\simeq\; \mathcal{O}(1) \, {\rm [GeV]},
\end{align}
for the dark sector Yukawa coupling $\Sigma \simeq 0.1$. 
At the same time, recalling the light neutrino masses formula in Eq.~\eqref{eq:mnu}, 
the suppressed mass $\epsilon \delta v_S = \mathcal{O}(1)$~[eV] 
is also consistent with experimental observations as well as the scale separation given in Eq.\eqref{eq:key}. 
Thus, both the realistic neutrino mass scale and the expected asymmetric DM mass are successfully accommodated within this framework.  

For later convenience, we denote the relevant Yukawa interaction terms $\mathcal{L}_Y$ allowed by the $U(1)_X$ in the mass basis as (neglecting $\mathcal{O}(1/\Lambda)$ suppression) 
\begin{align}\label{eq:Y3}
-\mathcal{L}_Y = \bar{\ell}_{i} Y_{ij} \tilde{H} P_R \nu^h_j 
+ f_a \, \eta \, \bar{\chi} P_R \psi_a 
+ \tilde{f}_a \, \eta^* \, \bar{\psi}_a P_R \chi 
+ g_i \, \phi \, \bar{\chi} P_R \nu^h_i 
+ \mathrm{h.c.},
\end{align}
where $P_R$ is the right-handed chiral projection operator,  
and $\nu^h_i$ denotes the heavy neutrinos with masses of order $\mathcal{O}(v_S)$.

We also note that, due to different charge assignments of the chiral fermions, 
$\chi_L$ couples to heavy neutrinos at leading order, while it does not directly couple to the light right-handed neutrinos.  
The unsuppressed interaction term $g_i \, \phi \, \bar{\chi} P_R \nu^h_i$ play an essential role 
in generating lepton and DM asymmetries via heavy neutrino decays in the early universe.  

Meanwhile, the Yukawa interactions between the scalar $\eta$ and $\chi_{L,R}$ or $\psi_{L,R}$ are crucial 
for depleting the symmetric thermal relic abundance of the DM, 
as well as for direct detection signals of DM through $\eta$- and Higgs-mediated processes, 
which will be discussed in Sec.~\ref{sec:pheno}.
We also stress that all unwanted interactions such as $\bar{\ell}_L H P_R \nu^\ell$ and $\phi \bar{\chi} P_R \nu^\ell$ 
are naturally suppressed by the Froggatt-Nielsen mechanism, whose 
processes will be discussed later.

\section{Leptogenesis and Asymmetric Dark Matter}\label{sec:leptogenesis}

We consider thermal leptogenesis via the decay of heavy Dirac neutrinos. 
The heavy states $\nu^h$, with masses $M_i \sim v_S \gg v_{\text{EW}}$, 
can decay into SM leptons through the Yukawa interactions in Eq.~\eqref{eq:Y3}. 
In a purely Dirac framework, the net lepton number asymmetry would vanish. 
However, in our model the global lepton number $U(1)_L$ is shared between the SM and the dark sector. 
This opens the possibility for simultaneous generation of both lepton and dark matter (DM) asymmetries 
via the CP-violating decays of $\nu^h$.

\begin{figure}[thbp]
  \centering
   \begin{tikzpicture}[baseline=(o.base)]
      \begin{feynhand}
         \vertex (a) at (-1.5,0) {$\nu^h$};
         \vertex (b) at (1,-1) {$H$};
         \vertex (c) at (1,1) {$\ell_L$};
         \vertex [dot] (o) at (0,0);
         \propag [fermion] (a) to (o);
         \propag [fermion] (o) to (c);
         \propag [scalar] (o) to (b);
      \end{feynhand}
   \end{tikzpicture}
\quad \quad \quad \quad 
   \begin{tikzpicture}[baseline=(o.base)]
      \begin{feynhand}
         \vertex (a) at (-1.5,0) {$\nu^h$};
         \vertex (b) at (1,-1) {$\phi^*$};
         \vertex (c) at (1,1) {$\chi$};
         \vertex [dot] (o) at (0,0);
         \propag [fermion] (a) to (o);
         \propag [fermion] (o) to (c);
         \propag [scalar] (o) to (b);
      \end{feynhand}
   \end{tikzpicture}
  \caption{Heavy neutrino decay channels.}
  \label{fig:nu}
\end{figure}
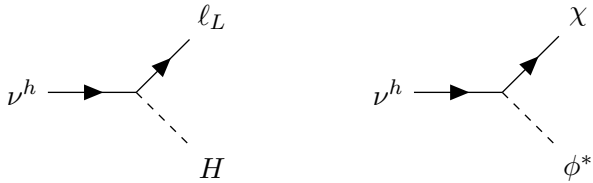

When the universe cools below the scale $v_S$, the heavy Dirac neutrinos fall out of equilibrium 
and decay into both the SM and dark sectors through two channels in Fig.~\ref{fig:nu}:
\[
\nu^h_i \to \ell_{L,j} H, 
\qquad 
\nu^h_i \to \chi \, \phi^*,
\]
with Yukawa couplings $Y_{ij}$ and $g_i$ defined in Eq.~\eqref{eq:Y3}.
Since $M_i \gg v_{\text{EW}}, m_\chi$, the dynamics can be studied in the unbroken electroweak phase, 
where SM fields are effectively massless, while $\nu^h_i$ and $\phi$ are heavy. 
At tree level, the total decay rate of $\nu^h_i$ is
\begin{align}
\Gamma_{\nu^h_i} &=
\left(
2 \sum_{j=1}^3 |Y_{ji}|^2 
+ \left(1 - \frac{m_\phi^2}{M_i^2}\right)^2 |g_i|^2
\right) 
\frac{M_i}{32\pi},
\label{eq:Gamma}
\end{align}
assuming $m_\phi < M_i$ for kinematic accessibility, consistent with the hierarchy in Eq.~\eqref{eq:hierarchy}.

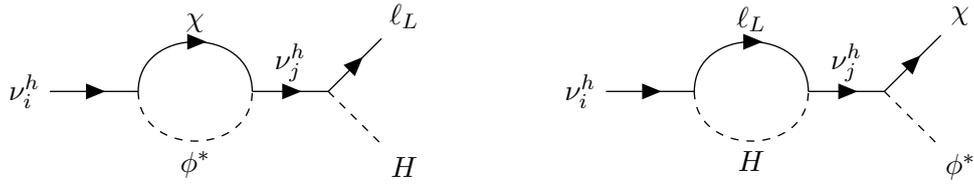
\begin{figure}[thbp]
  \centering
  \begin{tikzpicture}[baseline=(o.base)]
     \begin{feynhand}
         \vertex (a) at (-1.5,0) {$\nu^h_i$};
         \vertex (b) at (3.5,-1) {$H$};
         \vertex (c) at (3.5,1) {$\ell_L$};
         \vertex [dot] (o1) at (0,0);
         \vertex [dot] (o2) at (1.5,0);
         \vertex [dot] (o3) at (2.5,0);
         \propag [fermion] (a) to (o1);
         \propag[scalar] (o1) to [edge label' = $\phi^*$][out=270, in=270, looseness=1.5] (o2);
         \propag[fermion] (o1) to [edge label = $\chi$][out=90, in=90, looseness=1.5] (o2);
         \propag [fermion] (o2) to [edge label= $\nu^h_j$] (o3);
         \propag [scalar] (o3) to (b);
         \propag [fermion] (o3) to (c);
  \end{feynhand}
  \end{tikzpicture}
\quad \quad \quad \quad 
   \begin{tikzpicture}[baseline=(o.base)]
     \begin{feynhand}
         \vertex (a) at (-1.5,0) {$\nu^h_i$};
         \vertex (b) at (3.5,-1) {$\phi^*$};
         \vertex (c) at (3.5,1) {$\chi$};
         \vertex [dot] (o1) at (0,0);
         \vertex [dot] (o2) at (1.5,0);
         \vertex [dot] (o3) at (2.5,0);
         \propag [fermion] (a) to (o1);
         \propag[scalar] (o1) to [edge label' = $H$][out=270, in=270, looseness=1.5] (o2);
         \propag[fermion] (o1) to [edge label = $\ell_L$][out=90, in=90, looseness=1.5] (o2);
         \propag [fermion] (o2) to [edge label= $\nu^h_j$] (o3);
         \propag [scalar] (o3) to (b);
         \propag [fermion] (o3) to (c);
  \end{feynhand}
   \end{tikzpicture}
  \caption{One-loop diagrams for wave function corrections contributing to lepton and dark matter number asymmetries in the heavy Dirac neutrino decay channels.}
  \label{fig:nu1}
\end{figure}

The asymmetries are dominantly generated by the decay of the lightest heavy neutrino $\nu^h_1$. 
Although the total decay rates of $\nu^h_1$ and $\bar{\nu}^h_1$ are identical by CPT invariance, 
CP violation induces different branching fractions into the SM and dark sectors. 
We define the asymmetry parameters
\begin{align}
\epsilon_{\nu^h_1 \to \ell H} &=
{\rm Br}(\nu^h_1 \to \ell_L H) - {\rm Br}(\bar{\nu}^h_1 \to \bar{\ell}_L H^\dagger),
\nonumber \\
\epsilon_{\nu^h_1 \to \chi \phi^*} &=
{\rm Br}(\nu^h_1 \to \chi \phi^*) - {\rm Br}(\bar{\nu}^h_1 \to \bar{\chi} \phi),
\label{eq:epsilon}
\end{align}
which are calculated in one-loop order as shown in Fig.~\ref{fig:nu1}, 
\begin{align}
\epsilon_{\nu^h_1 \to \ell H} = - \epsilon_{\nu^h_1 \to \chi \phi^*}
&= 
\frac{1}{16\pi} 
\frac{ \sum_{k\neq 1} \frac{4 M_1^2}{M_1^2 - M_k^2} 
\, {\rm Im}\!\left[g_1^* (Y^\dagger Y)_{1k} g_k \right] }
{ \sum_{i=1}^3 2|Y_{i1}|^2 + \left(1 - \frac{m_\phi^2}{M_1^2}\right)^2 |g_1|^2 }
\left(1 - \frac{m_\phi^2}{M_1^2}\right)^2 .
\end{align}
A nonzero asymmetry thus requires at least two generations of heavy neutrinos.

\begin{figure}[thbp]
  \centering
  \begin{tikzpicture}[baseline=(o.base)]
     \begin{feynhand}
         \vertex (a) at (-2,1) {$\phi, \phi^*$};
         \vertex (b) at (-2,-1) {$\phi, \phi^*$};
         \vertex (c) at (2,0) {$\eta_I$};
         \vertex [dot] (o1) at (0,0);
         \propag [scalar] (a) to (o1);
         \propag [scalar] (o1) to (b);
         \propag [scalar] (o1) to (c);
  \end{feynhand}
  \end{tikzpicture}
  \caption{  Feynman diagram for the complex scalar $\phi$ or $\phi^*$ pair annihilation process into $\eta_I$.}
  \label{fig:phi}
\end{figure}
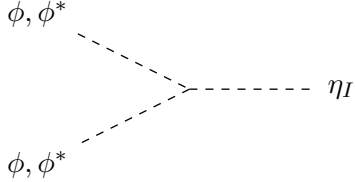

Moreover, the simultaneously generated number asymmetry for the complex scalar $\phi$ is efficiently washed out: 
A pair of $\phi$ rapidly annihilates via the trilinear scalar interaction $\lambda_{S\phi\eta} v_S \phi^2 \eta^* + {\rm h.c.}$ 
from the potential $V$ in Eq.~\eqref{eq:V}, as illustrated in Fig.~\ref{fig:phi}.
The processes that violate the $\phi$ number remain in thermal equilibrium throughout the relevant epoch.
In the relativistic regime $T \simeq m_\phi  < v_S$, the $s$-channel annihilation cross section mediated by $\eta_I$ is estimated as
\begin{align}\label{eq:annihilation}
\langle \sigma v_{\rm rel}\rangle_{\phi \phi \to \phi^* \phi^*} = \frac{|\lambda_{S\phi\eta}|^4}{256\pi} \left(\frac{v_S}{T}\right)^4 \frac{1}{T^2}.
\end{align}
Using the number density $n_\phi = \dfrac{\zeta(3)}{\pi^2} T^3$, the interaction rate
$\Gamma_{\phi\phi \to \phi^*\phi^*} = n_\phi \langle \sigma v_{\mathrm{rel}} \rangle$
can be compared with the Hubble parameter $H = 1.66 g_\ast^{1/2} T^2 / M_{\mathrm{Pl}}$:
\begin{align}
\frac{\Gamma_{\phi\phi\to\phi^*\phi^*}}{H} \simeq 10^{-5}  |\lambda_{S\phi\eta}|^4 \left(\frac{v_S}{T}\right)^4 \frac{M_{\rm Pl}}{T} \gg 1 
\end{align}
for $v_S \simeq 10^{10}~\text{[GeV]}$ and $\lambda_{S\phi\eta} = \mathcal{O}(1)$.
Thus this annihilation channel remains in thermal equilibrium during the decay of the heavy Dirac neutrinos, 
and any $\phi$ number asymmetry is efficiently washed out. 
In addition, the symmetric component of $\phi$ can further annihilate into lighter scalars through both two-body and three-body channels such as
$\phi\phi^{*} \to \eta_I \eta_I$, $\phi\phi(\phi^*\phi^*) \to \eta_I \eta_I \eta_I$,
mediated by the quartic couplings $\lambda_\eta, \lambda_{S\phi\eta}$, and $\lambda_{\phi\eta}$. 
With $\mathcal{O}(1)$ values of these couplings and a large mass hierarchy $m_\phi \gg m_{\eta_I}$, 
the annihilation cross sections are sufficiently enhanced to ensure that the relic density of $\phi$ becomes negligible, leaving no impact on the DM abundance.

Once a lepton (or $B-L$) asymmetry is generated from the heavy neutrino decays, 
part of this lepton asymmetry is converted into a baryon asymmetry through 
the $B+L$ violating sphaleron process~\cite{Kuzmin:1985mm}, 
which is active above the critical temperature of the electroweak phase transition $T_c \leq 150~{\rm [GeV]}$.
If we demand that $B-L = B-L_{\rm SM} - D = 0$, 
a net baryon asymmetry $B \neq 0$ is induced from a nonzero asymmetry $B-L_{\rm SM} \neq 0$, 
where $D$ denotes the total DM number carried by $\chi$ or $\psi$, 
and $L_{\rm SM}$ is the lepton number of SM particles excluding the right-handed neutrinos.

A salient feature of our framework is that, unlike in conventional Dirac leptogenesis (or neutrinogenesis),
the generated asymmetry does not reside in the right-handed neutrino sector but instead in the right-handed asymmetric dark matter sector.
Because total lepton number is exactly conserved, no net asymmetry is produced; rather, equal and opposite asymmetries arise in the dark and SM sectors.
Consequently, the dark and visible asymmetries are tightly correlated.
In our setup, the dark-sector asymmetry is carried by the field $\chi$, which is stabilized by its additional $\mathbb{Z}_4^D$ charge.

For this mechanism to operate consistently, the dark and SM sectors must not be in chemical equilibrium.
More precisely, interactions that change the DM number must be frozen out.
To demonstrate this, we examine the effective Yukawa couplings that mediate interactions between the DM and lepton sectors.
The couplings corresponding to the operators $\bar{\chi}_L \nu_R$ and $\bar{\psi} \nu_R$, 
which could otherwise transfer a portion of the dark asymmetry into a $\nu_R$ asymmetry, are suppressed by a factor of $\delta$.
As a result, the cross sections for DM-number-changing processes such as
$\chi\psi \to \nu^\ell \nu^\ell$ and $\chi\chi \to \nu^\ell \nu^\ell \eta_I$,
which are allowed by the $\mathbb{Z}_4^D$ symmetry, are suppressed by at least $\mathcal{O}(\delta^4)$.
Similarly, the effective Yukawa couplings for the charged leptons and light neutrinos arise from
\[
\, \bar{\ell}_{L,i} Y_{ij}' \tilde{H} \, N_R 
\quad \quad  {\rm or} \quad \quad 
\frac{S^*}{\Lambda} \, \bar{\ell}_{L,i} \, \tilde{Y}'_{ij} \tilde{H} \, \nu_{R,j}, 
\]
which are also suppressed to $\mathcal{O}(\delta)$ due to the tiny mixing in the Dirac seesaw mass matrix (see Eq.~\eqref{eq:VLR}) 
or by the higher-dimensional suppression associated with the $U(1)_X$ symmetry.
These interactions can induce DM-number-changing processes of the form
$\chi\phi^* \to \ell_L H$.
All relevant tree-level diagrams are shown in Fig.~\ref{fig:conv}.
Among them, the process $\chi\phi^* \to \ell_L H$ provides the dominant contribution,
as the other two channels are further suppressed by factors of $1/m_{\phi}^2$ due to the heavy $\phi$ propagators.
To estimate the magnitude of the conversion rate, we evaluate the cross section in the relativistic limit.
For the $s$-channel exchange mediated by $\nu^\ell$ at temperature $T$, we obtain
\begin{align}\label{eq:conversion}
\langle \sigma v_{\rm rel}\rangle_{\chi \phi^* \to \ell_{L,i} H} = 
\sum_{j=1}^3 \frac{1}{32\pi} \delta^4 |h'_{j}|^2 |\tilde{Y}'_{ij}|^2 \frac{m_\chi m_{\ell_i}}{T^2} \frac{1}{T^2},
\end{align}
where $m_{\ell_i}$ is the mass of the lepton $\ell_i$,
and $h'_j$ denotes the coupling between $\chi$ and $\nu_R$ defined in Eq.~\eqref{eq:Y2}.
The corresponding ratio of the interaction rate to the Hubble parameter is then estimated as
\begin{align}
\frac{\Gamma_{\chi \phi^* \to \ell_{L,i} H}}{H} \simeq 10^{-3}  \delta^4 
|h'_{j}|^2 |\tilde{Y}'_{ij}|^2 \frac{m_\chi m_{\ell_i}}{T^2} \frac{M_{\rm Pl}}{T} \ll 1, 
\end{align}
where the extreme suppression follows from $\delta \simeq 10^{-9}$.
Thus, DM-number-changing processes mediated by these interactions remain far out of equilibrium throughout the relevant epoch.
In addition, as further examined in the DM direct detection analysis of Sec.~\ref{sec:pheno},
the annihilation amplitudes of a $\chi+\bar{\chi}$ pair into SM particles are also strongly suppressed due to the Higgs-portal structure of our setup.
For representative values $\lambda_{H\eta} \sim 10^{-3}$ and $h_a, \tilde{h}_a \sim 0.1$, 
which are consistent with direct detection constraints discussed in Sec.~\ref{sec:pheno},
the corresponding DM pair annihilation processes never reach thermal equilibrium.
Therefore, the dark and visible sectors remain chemically decoupled throughout the entire thermal history of the universe.
Moreover, these parameter choices automatically satisfy the standard out-of-equilibrium conditions required in conventional Dirac leptogenesis~\cite{Dick:1999je}.
In particular, the smallness of the effective Yukawa couplings ensures that the lepton asymmetry does not equilibrate between 
$\nu_L$ and $\nu_R$ before the electroweak phase transition, 
and also prevents thermalization of right-handed neutrinos with SM particles during leptogenesis.

\begin{figure}[thbp]
\centering
\begin{tikzpicture}
\begin{feynhand}
         \vertex (ul) at (-1,1) {$\chi$};
         \vertex (dl) at (-1,-1) {$\phi^*$};
	 \vertex (ur) at (2.5,1) {$\ell_L$};
	 \vertex (dr) at (2.5,-1) {$H$};
         \vertex [dot] (o1) at (0,0);
         \vertex [dot] (o2) at (1.5,0);
         \propag [fermion] (ul) to (o1);
         \propag [scalar] (dl) to (o1);
         \propag [fermion] (o2) to (ur);
         \propag [scalar] (o2) to (dr);
         \propag [fermion] (o1) to [edge label = $\nu^\ell$] (o2);
         \vertex (ul) at (3.5,1.) {$\chi$};
         \vertex (dl) at (3.5,-1.) {$\psi$};
	 \vertex (ur) at (6.5,1.) {$\nu^\ell$};
	 \vertex (dr) at (6.5,-1.) {$\nu^\ell$};
         \vertex [dot] (o1) at (5,0.5);
         \vertex [dot] (o2) at (5,-0.5);
         \propag [fermion] (ul) to (o1);
         \propag [fermion] (dl) to (o2);
         \propag [fermion] (o1) to (ur);
         \propag [fermion] (o2) to (dr);
         \propag [scalar] (o1) to [edge label = $\phi$] (o2);
         \vertex (ul2) at (7.5,1) {$\chi$};
         \vertex (dl2) at (7.5,-1) {$\chi$};
	 \vertex (ur2) at (12,2) {$\nu^\ell$};
	 \vertex (dr2) at (12,-2) {$\nu^\ell$};
	 \vertex (c2) at (12,0) {$\eta_{I}$};
         \vertex [dot] (o12) at (9,1);
         \vertex [dot] (o22) at (9,-1);
         \vertex [dot] (o32) at (10.5,0);
         \propag [fermion] (ul2) to (o12);
         \propag [fermion] (o12) to (ur2);
         \propag [fermion] (dl2) to (o22);
         \propag [fermion] (o22) to (dr2);
         \propag [scalar] (o12) to [edge label = $\phi$] (o32);
         \propag [scalar] (o22) to [edge label' = $\phi$] (o32);
         \propag [scalar] (o32) to (c2);
\end{feynhand}
\end{tikzpicture}
\caption{Feynman diagrams for DM number conversion processes.}
\label{fig:conv}
\end{figure}

Since sphalerons do not act on $\chi$ or $\psi$ (both being $SU(2)_L$ singlets), 
at temperatures near and above $T_c$ the equilibrium conditions among SM species 
lead to the usual sphaleron relation between baryon and lepton asymmetries~\cite{Harvey:1990qw}. 
Neglecting $\nu_L^\ell$-$\nu_R^\ell$ equilibration, one obtains
\begin{align}
n_B = \frac{28}{79}\, n_{B-L_{\rm SM}}, 
\label{eq:nB}
\end{align}
where $n_B$ is the baryon number density.
The DM number density $n_\chi$ is related to $n_B$ via
\begin{align}
n_\chi = \frac{79}{28} \, n_B.  
\label{eq:n_chi}
\end{align}
The relic asymmetric DM density is then
\begin{align}
\Omega_{\rm asy} = \frac{n_\chi m_\chi}{\rho_c},
\label{eq:Omega}
\end{align}
with $\rho_c$ the critical density. 
From cosmology, the observed ratio is $\Omega_{\rm DM}/\Omega_{B} \simeq 5$, 
with $\Omega_{\rm DM} \simeq 0.26$. 
Assuming $\Omega_{\rm DM} = \Omega_{\rm asy}$, one finds
\begin{align}
\frac{\Omega_{\rm DM}}{\Omega_{B}} 
= \frac{n_\chi m_\chi}{n_B m_p} 
= \frac{79}{28}\, \frac{m_\chi}{m_p}, 
\end{align}
where $m_p$ is the nucleon mass, so that the DM mass is expected to be $m_\chi = 1.8~{\rm [GeV]}$.
As will be discussed, 
since the relic symmetric component of DM energy density ($\Omega_{\chi}$) can contribute to 
total DM density, for a general case the dark matter mass can be estimated as 
\begin{align}\label{eq:mx}
m_\chi = \left(5-\frac{\Omega_{\chi}}{\Omega_{B}}\right) \frac{28}{79}\, m_p. 
\end{align}

\subsection*{Out-of-equilibrium condition}

In addition to CP violation, successful leptogenesis requires out-of-equilibrium decays. 
For the lightest heavy neutrino $\nu^h_1$, we define the decay parameter as 
\begin{align}
K_1 \equiv \left. \frac{\Gamma_{\nu^h_1}}{H}\right|_{T=M_1},
\end{align}
where $H$ is the Hubble parameter, 
and $g_\ast$ denotes the effective number of relativistic degrees of freedom~\cite{Kolb:1990vq}. 
To account for the smallness of active neutrino masses, we require 
$M_1 \simeq v_S \gtrsim 10^{10}~{\rm [GeV]}$, 
so that leptogenesis typically occurs in the strong washout regime ($K_1 \gg 1$ for $Y_{ij}, g_i \leq \mathcal{O}(1)$). 
Solving the relevant Boltzmann equations, 
we can obtain the final amount of the baryon-to-photon ratio as 
\begin{align}
\eta_B \simeq 0.01 \, \kappa \, \epsilon_{\nu^h_1 \to \ell H},   
\end{align}
where $\kappa$ is the efficiency factor, which arises due to inverse decay and scattering processes~\cite{Buchmuller:2004nz}.
For $K_1 \ll 1$, the efficiency factor $\kappa$ is close to unity, while for $K_1 \gg 1$ it is suppressed as an inverse power of $K_1$~\cite{Buchmuller:2004nz,Cerdeno:2006ha}. 
Unlike standard Majorana leptogenesis, 
the CP asymmetry here depends both on the dark sector couplings $g_i$ and the Higgs Yukawa matrix $Y$. 
Since $Y$ is not tied directly to light neutrino masses, no Davidson-Ibarra bound applies~\cite{Davidson:2002qv}, 
while in our setup, with $v_S \simeq 10^{10}~{\rm [GeV]}$, this requirement is already satisfied.
For numerical analysis we simply choose  
$\displaystyle \kappa = \frac{0.12}{K_1^{1.1}}$ 
for $K_1 >1$ and $\displaystyle \kappa = 1$ for $K_1<1$ following~\cite{Borboruah:2024lli}. 
The resulting value can be compared with the observed $\eta_B = 6.1 \times 10^{-10}$~\cite{Planck:2018vyg}.

\subsection*{Numerical Analysis}

We perform a numerical scan over the parameter space defined by 
$10^{-4} < |Y_{ij}| < 0.01$ and $10^{-4} < |g_i| < 1$, 
both uniformly distributed on a logarithmic scale, 
with complex phases in the ranges 
$-\pi < \arg(Y_{ij}) < \pi$ and $-\pi < \arg(g_i) < \pi$ for $i,j = 1,3$. 
The upper bound $|Y_{ij}| < 0.01$ is consistent with the expected magnitude 
required to reproduce the observed light neutrino masses. 
The heavy neutrino mass hierarchy is fixed as $M_2 = 1.5\,M_1$ and $M_3 = 2\,M_1$, 
while we set $m_\phi = M_1/10$ as a representative choice. 

\begin{figure}[htbp]
\centering
\includegraphics[width=0.45\textwidth]{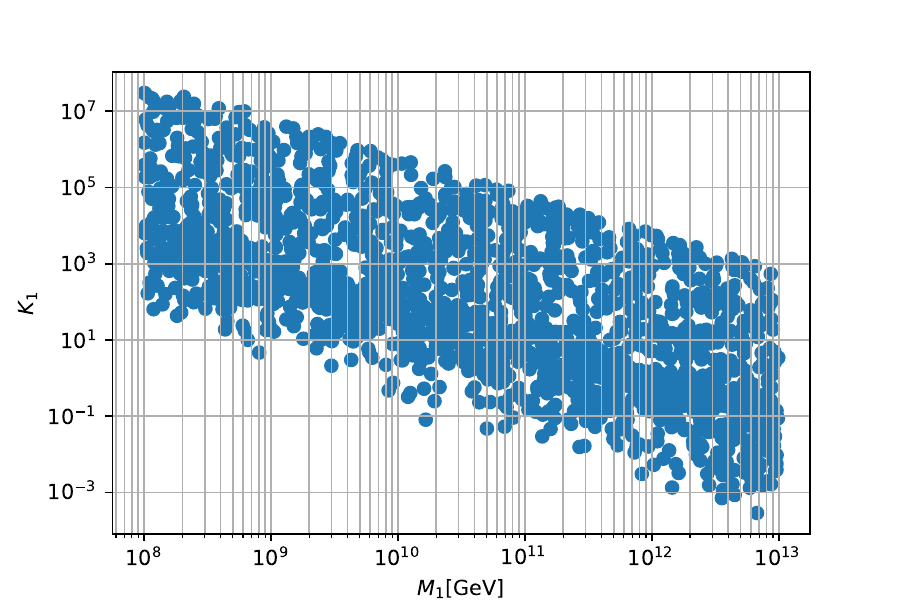} 
\quad 
\includegraphics[width=0.45\textwidth]{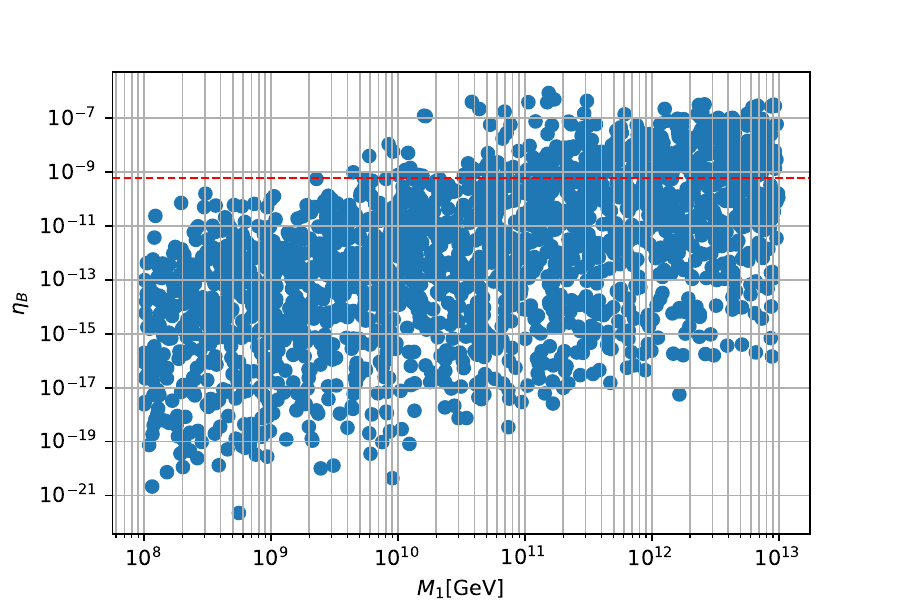} 
\caption{
Left: Dependence of the decay parameter $K_1$ on $M_1$.
Right: Resulting baryon-to-photon ratio $\eta_B$ as a function of $M_1$.
The horizontal red dashed line indicates the observed value of $\eta_B$.}
\label{fig:eta_B}
\end{figure}

Figure~\ref{fig:eta_B} shows the dependence of the decay parameter $K_1$ (left) 
and the resulting baryon-to-photon ratio $\eta_B$ (right) on $M_1$. 
The observed baryon asymmetry can be successfully reproduced for $M_1 \gtrsim 10^{10}\,\mathrm{[GeV]}$,
which is consistent with our assumption of $v_S \simeq 10^{10}\,\mathrm{[GeV]}$. 
The dependence on $m_\phi$ is negligible as long as $m_\phi$ is smaller than $M_1$, 
since $(1 - m_\phi^2/M_1^2)^2 \simeq 1$ in this limit.

\section{Cosmological Constraints}\label{sec:pheno}

In our scenario, a substantial relic density remains from the asymmetric component of $\chi$, whose contribution depends on its mass $m_\chi$. 
However, the relic density from the symmetric component must also be considered, as it contributes to the total DM abundance and must be consistent with cosmological observations. 

We thus have two DM candidates, $\chi$ and $\psi$, both carrying the same lepton number and non-trivial $\mathbb{Z}_4^D$ charges. 
At tree level, $\psi$ and $\chi$ can interact through the Yukawa interactions mediated by the scalar $\eta_I$. 
An important case arises when 
\begin{align}
m_{\eta_I} < m_\chi \ (< m_\psi),
\end{align}
for which the symmetric component of dark matter can dominantly annihilate into a pair of light scalars $\eta_I$. 
This scenario also opens the possibility of probing DM through direct detection experiments through Higgs-mediated nucleon scattering. 
In this context, the portal coupling $\lambda_{H\eta}$ plays a crucial role, as it provides the only connection between the SM and dark sectors. 

The real scalar $\eta_I$ attains thermal equilibrium with the SM bath through elastic scatterings of the form $\eta_I X \to \eta_I X$, 
where $X$ denotes SM fermions, gauge bosons, or the Higgs boson. If $\eta_I$ remains in kinetic equilibrium with the SM bath until the dark matter freeze-out temperature $T_f$, 
the relic abundance can be computed analogously to the conventional WIMP framework~\cite{Rubakov:2017xzr,Gondolo:2012vh}. 
This consideration accommodates a viable scenario with a light MeV-scale mediator satisfying
\begin{align}
T_{\rm BBN} < m_{\eta_I} < m_\chi = \mathcal{O}(1)\,{\rm [GeV]}.
\end{align}
Cosmological observations, particularly those from Big Bang Nucleosynthesis (BBN) and the Cosmic Microwave Background (CMB) power spectrum, impose stringent bounds on additional relativistic species that remain in thermal contact with the SM during the BBN epoch, corresponding to $T_{\rm BBN} \simeq 1\,{\rm [MeV]}$.

In the following, we investigate the cosmological constraints associated with the asymmetric DM candidate $\chi$. 

\subsection*{Thermal Relic Abundance of the DM}

\begin{figure}[thbp]
  \centering
  \begin{tikzpicture}[baseline=(o.base)]
     \begin{feynhand}
         \vertex (a) at (-2,1) {$\chi$};
         \vertex (b) at (-2,-1) {$\bar{\chi}$};
         \vertex (c) at (2,1) {$\eta_I$};
         \vertex (d) at (2,-1) {$\eta_I$};
         \vertex [dot] (o1) at (0,1);
         \vertex [dot] (o2) at (0,-1);
         \propag [fermion] (a) to (o1);
         \propag [fermion] (o2) to (b);
         \propag [scalar] (o1) to (c);
         \propag [scalar] (d) to (o2);
         \propag [fermion] (o1) to [edge label' = $\psi_a$] (o2);
  \end{feynhand}
  \end{tikzpicture}
  \caption{Dominant Feynman diagram for dark matter pair annihilation process.}
  \label{fig:DM}
\end{figure}
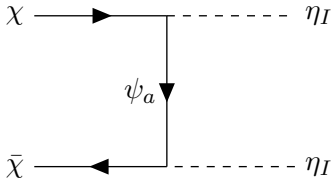

For $m_{\eta_I} < m_\chi = \mathcal{O}({\rm [GeV]})$, 
the relic abundance of the symmetric component of $\chi$ is determined by its annihilation cross section, as illustrated in Fig.~\ref{fig:DM}. 
The thermally averaged annihilation cross section in the non-relativistic limit is given by
\begin{align}\label{eq:sigma_ann}
\langle \sigma_{\rm ann} v_{\rm rel}\rangle_{\chi \bar{\chi} \to \eta_I \eta_I} = 
\frac{A}{64\pi m_\chi^2}\sqrt{1-\frac{m_{\eta_I}^2}{m_\chi^2}},
\end{align}
where
\begin{align}\notag
A =& \ \frac{m_\chi^2}{2} \sum_{a,b=1}^4 
\big[ (|f_a|^2+|\tilde{f}_a|^2)(|f_b|^2+|\tilde{f}_b|^2)(m_\chi^2-m_{\eta_I}^2) 
+ 4\, {\rm Im}( f_a \tilde{f}_a)\,{\rm Im}( f_b \tilde{f}_b)\, m_{\psi_a} m_{\psi_b} \big] 
\\
&\hspace{15mm} \times 
\frac{1}{m_{\eta_I}^2-m_\chi^2-m_{\psi_a}^2}\,
\frac{1}{m_{\eta_I}^2-m_\chi^2-m_{\psi_b}^2}.
\end{align}

Assuming $\eta_I$ remains in kinetic equilibrium with the SM particles, 
the relic density of the symmetric component of $\chi$ after freeze-out is reasonably estimated as
\begin{align}
\Omega_{\chi}h^2 \simeq g \times 
\frac{1.07\times 10^9 \, x_f}{(g_{*s}/g_*^{1/2})\, M_{\rm Pl}\, {\rm [GeV]}\, \langle \sigma_{\rm ann} v_{\rm rel} \rangle},
\end{align}
where $g$ is the internal degrees of freedom of $\chi$, $g_*$ denotes the effective relativistic degrees of freedom at freeze-out temperature $T=T_f$~\cite{Kolb:1990vq}.
The freeze-out parameter $x_f = m_\chi/T_f$ is determined by
\begin{align}
x_f = \frac{m_\chi}{T_f} &= 
\log\!\left[0.038 \left(\frac{g}{g_*^{1/2}}\right) M_{\rm Pl} m_\chi \langle \sigma_{\rm ann} v_{\rm rel} \rangle \right] 
- \frac{1}{2}\log\!\left\{ \log\!\left[0.038 \left(\frac{g}{g_*^{1/2}}\right) M_{\rm Pl} m_\chi \langle \sigma_{\rm ann} v_{\rm rel} \rangle \right]\right\}.
\end{align}
with $g_{*s}$ the effective relativistic degrees of freedom for entropy.

\begin{figure}[htbp]
\centering
\includegraphics[width=0.45\textwidth]{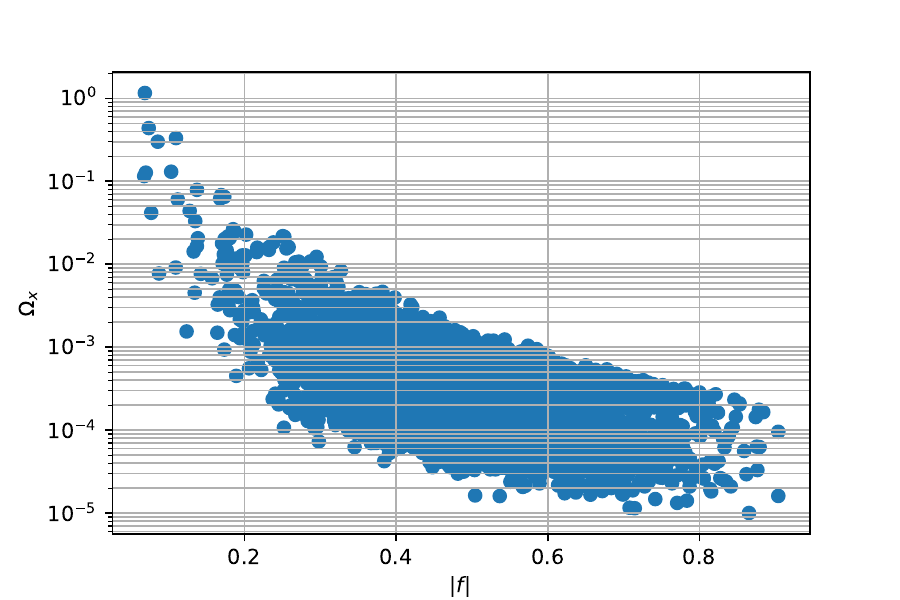} 
\quad 
\includegraphics[width=0.45\textwidth]{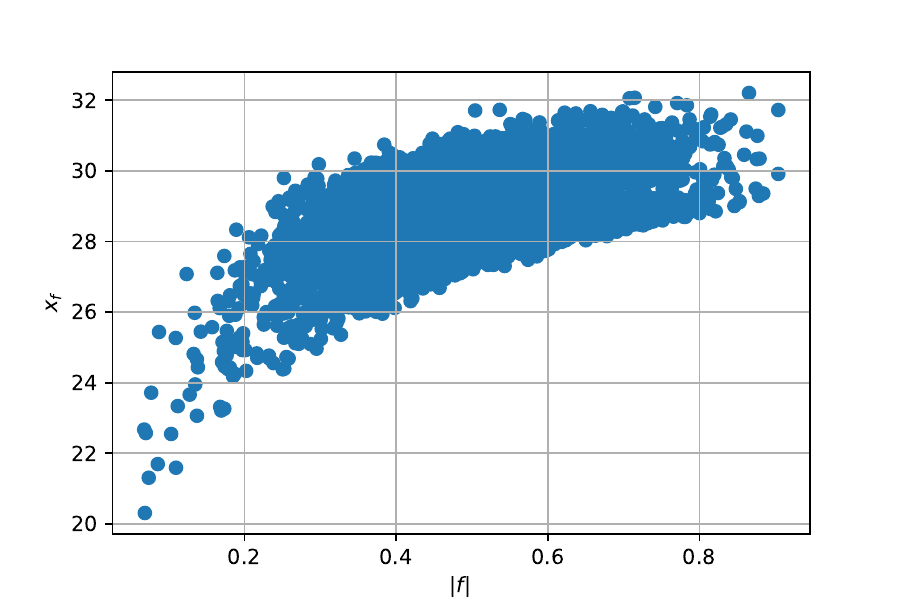} 
\caption{
Thermal relic abundance of the symmetric component of DM. 
The left panel shows the dependence on Yukawa couplings, and the right panel shows the corresponding freeze-out parameter $x_f$.}
\label{fig:Omega}
\end{figure}

\begin{figure}[htbp]
\centering
\includegraphics[width=0.45\textwidth]{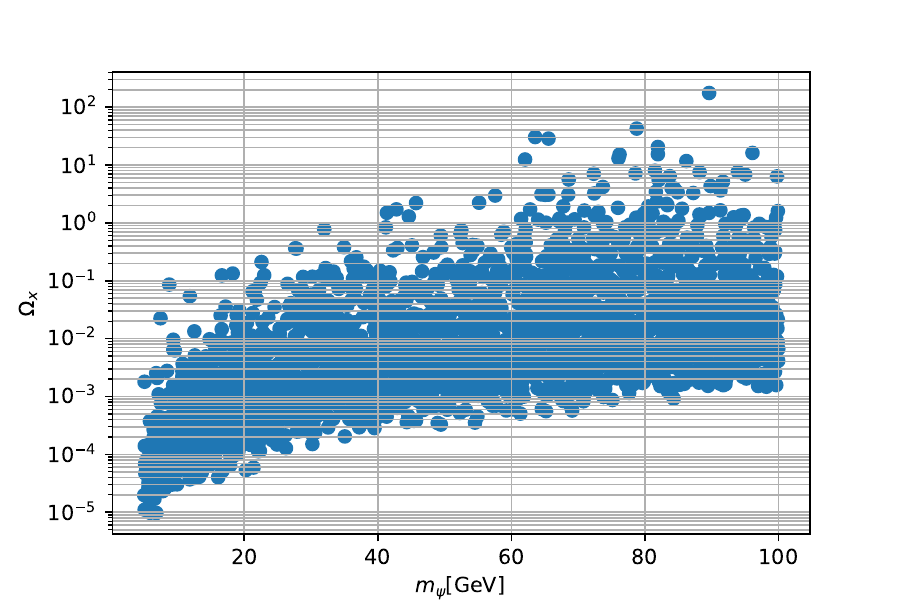} 
\caption{
Thermal relic abundance of the symmetric component of DM as a function of $m_\psi$.}
\label{fig:Omega_mpsi}
\end{figure}

Numerical results are shown in Figs.~\ref{fig:Omega} and \ref{fig:Omega_mpsi}, 
where we assume a common absolute value of the Yukawa couplings, independent of flavor, as 
$|f_a| = |\tilde{f}_a| \equiv |f|$ for $a=1,\cdots,4$ with $0 < |f| < 1$, 
in order to clearly illustrate the impact of the couplings. 
The complex phases are randomly chosen in the range $-\pi < \arg(f_{a}), \arg(\tilde{f}_{a}) < \pi$. 
The masses of $\psi_a$, which must be heavier than $\chi$, are taken to be flavor-universal, 
$m_{\psi_{1,2,3,4}} \equiv m_\psi$, for simplicity. 
The mass parameter $m_\psi$ also plays a substantial role in determining the thermal relic abundance, 
while the dependence on $m_{\eta_I}$ is found to be small within the simulated region $m_{\eta_I} < m_\chi/2$.

In the left panel of Fig.~\ref{fig:Omega}, 
we fix $m_\chi =1.8\,{\rm [GeV]}$ and $m_\psi = 10\,{\rm [GeV]}$ and vary $5~[{\rm MeV}] < m_{\eta_I} < m_\chi/2$. 
The results show that the DM candidate becomes effectively a fully asymmetric dark matter, 
when the additional Yukawa interactions among $\chi$, $\psi$, and $\eta_I$ are sufficiently strong 
($|f| \gtrsim 0.15$). 
In this case, the dark sector species remain in thermal equilibrium, sharing a common temperature. 
The corresponding values of $x_f$ are displayed in the right panel of the figure. 
We find $x_f \simeq 20\text{--}30$, which implies a freeze-out temperature 
$T_f \gtrsim 60\,{\rm [MeV]}$ for $m_\chi = 1.8$[GeV]. 
This safely avoids constraints from BBN.\footnote{
For more precise estimates of the thermal relic density, 
one should solve the coupled Boltzmann equations including the dark sector temperature. 
See, e.g., \cite{Dutta:2022knf} for a similar setup in which a quantitatively consistent result has been obtained.
See also \cite{Biswas:2021kio} for numerical analysis of the scalar decoupling temperature for similar $\lambda_{H\eta}$ values.}

Figure~\ref{fig:Omega_mpsi} illustrates the dependence of the relic abundance on the mass $m_\psi$. 
Here, we again assume degenerate masses $m_{\psi_{1,2,3,4}} \equiv m_\psi$, 
and vary $5~[{\rm MeV}] < m_{\eta_I} < m_\chi/2$.  
While $\Omega_\chi$ increases with $m_\psi$, 
a fully asymmetric DM scenario can still be realized over a wide range of $m_\psi$, 
depending on the values of the couplings $f_a$ and $|\tilde{f}_a|$. 

\subsection{DM Direct Detection}
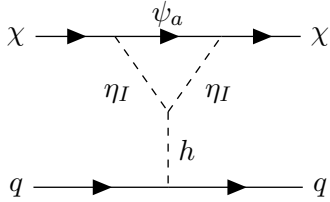
\begin{figure}[thbp]
  \centering
   \begin{tikzpicture}[baseline=(o.base)]
      \begin{feynhand}
         \vertex (a) at (-2,1) {$\chi$};
         \vertex (b) at (2,1) {$\chi$};
         \vertex (c) at (-2,-1) {$q$};
         \vertex (d) at (2,-1) {$q$};
         \vertex [dot] (o) at (0,0);
         \vertex [dot] (o1) at (-0.7,1);
         \vertex [dot] (o2) at (0.7,1);
         \vertex [dot] (o3) at (0,-1);
         \propag [fermion] (a) to (o1);
         \propag [fermion] (o1) to [edge label= $\psi_a$] (o2);
         \propag [fermion] (o2) to (b);
         \propag [scalar] (o1) to [edge label'=$\eta_I$]  (o);
         \propag [scalar] (o) to [edge label'= $\eta_I$]  (o2);
         \propag [scalar] (o) to [edge label= $h$]  (o3);
         \propag [fermion] (c) to (o3);
         \propag [fermion] (o3) to (d);
      \end{feynhand}
   \end{tikzpicture}
  \caption{Dark matter coupling to the quark.}
  \label{fig:SI}
\end{figure}

As $\eta$ does not acquire a vev, 
it does not mix with SM Higgs and hence the tree-level DM-nucleon scattering through the $H-\eta$ 
mixing portal is not possible. 
It can, however still scatter off the detector nucleus efficiently depending on the portal coupling $\lambda_{H\eta}$, 
the new Yukawa couplings $f_a, \tilde{f}_a$ and the masses $m_{\eta_I}, m_\psi$. 
The possibility of spin-independent(SI) DM nucleon elastic scattering allows for the detection of DM in terrestrial laboratories. 
Thus, in our case, the simplest diagram for direct detection is at the one-loop level, as depicted in Fig.~\ref{fig:SI},
where the real scalar $\eta_I$ and $\psi$ are running into the loop and the SM Higgs is playing the role of the mediator. 

The Higgs exchange diagram induces an effective scalar
interaction terms between the dark matter $\chi$ and the quark $q$ given as 
\begin{align}
\mathcal{L}_{\rm eff} 
= \sum_{q=u\sim t}
\frac{\lambda_{H\eta} m_q}{m_{h^0}^2} C_S \bar{\chi}\chi \bar{q} q.
\end{align}
$C_S$ is a quark-level effective coupling of the scalar interaction given as 
\begin{align}
C_S =& 
\frac{1}{2(4\pi)^2}
\sum_{a=1\sim4} 
\left( (f_a \tilde{f}_a + \tilde{f}_a^* f_a^*) m_{\psi_a} C_0(m_\chi^2) + m_{\psi_a} (|f_a|^2 + |\tilde{f}_a|^2) (C_0(m_\chi^2)+C_2(m_\chi^2)) \right),
\end{align}
where 
$C_0(p^2)$ and $C_2(p^2)$ are the loop functions defined as  
\begin{align}
\frac{i}{(4\pi)^2} C_0(p^2) \equiv&
\int\frac{d^4\ell}{(2\pi)^4}
\frac{1}{[(p+\ell)^2-m_{\psi_a}^2] \left[\ell^2-m_{\eta_I}^2\right]^2},   
\\ \notag
\frac{i}{(4\pi)^2} p^\mu C_2(p^2) \equiv& 
\int\frac{d^4\ell}{(2\pi)^4} 
\frac{\ell^\mu}{[(p+\ell)^2-m_{\psi_a}^2] \left[\ell^2-m_{\eta_I}^2\right]^2}.
\end{align}
The detailed expression for these functions and their derivations can be found in \cite{Ertas:2019dew}. 
See also \cite{Abe:2018emu} for slightly different setup of models, 
where a similar result for effective couplings can be obtained.
Using the effective couplings $C_S$, we obtain the dominant SI DM-nucleon ($N$) scattering cross section $\sigma^{\rm SI}_{N}$ given as 
\begin{align} \label{eq:sigma}
\sigma^{\rm SI}_{N} = \frac{\mu_N^2 |C_{\rm eff}^{\rm SI}|^2}{\pi}, 
\end{align}
where $\displaystyle \mu_N = \frac{m_\chi m_N}{m_\chi+m_N}$ with $m_N$ being the nucleon mass. 
$C_{\rm eff}^{\rm SI}$ is a DM-nucleon effective coupling which are given \cite{Ertas:2019dew}, 
\begin{align}\notag
C_{\rm eff}^{\rm SI} = 
\sum_{q=u,d,s} m_N f^N_q \frac{\lambda_{H\eta}}{m_{h^0}^2}C_S
+3 \frac{2}{27} m_N f_G^N \frac{\lambda_{H\eta}}{m_{h^0}^2}C_S,
\end{align}
with $f^N_q \lesssim 0.05$ and $f_G^N = 0.92$ are the nucleon scalar form factor (See also \cite{Hoferichter:2017olk} for a recent review of the DM direct detection). 

\begin{figure}[htbp]
\centering
\includegraphics[width=0.45\textwidth]{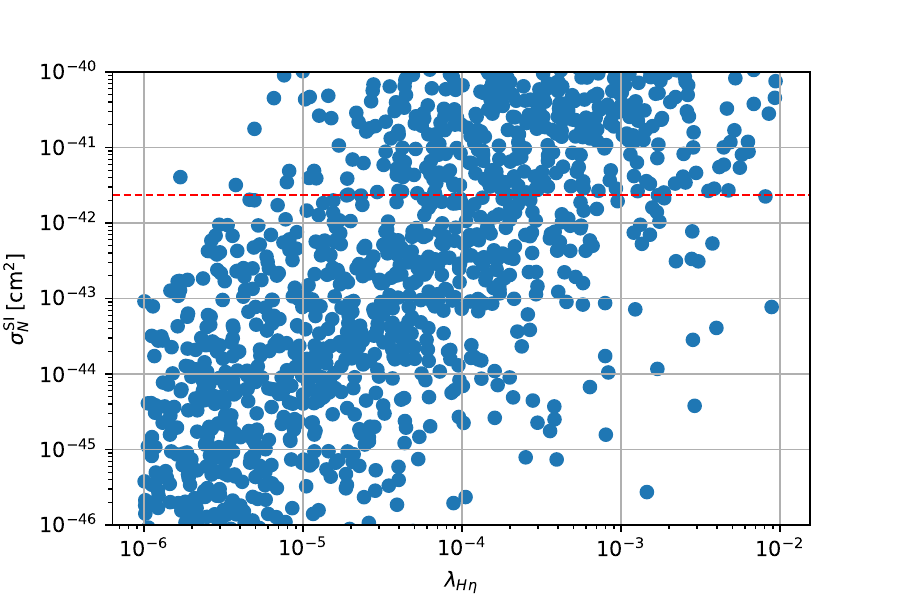} 
\caption{
Spin-Independent Nucleon-DM cross section for DM direct detection experiments. 
The horizontal line represents the current experimental bound for $m_\chi = 1.8$[GeV] (red dashed line) 
in the published 90\% C.L. limits from DS-50~\cite{DarkSide-50:2022qzh}.} 
\label{fig:sigma}
\end{figure}

Let us now numerically investigate the spin-independent elastic scattering cross section $\sigma^{\rm SI}_N$, defined in Eq.~\eqref{eq:sigma}.
As in the thermal relic analysis of Eq.~\eqref{eq:sigma_ann}, the quantities $\sigma^{\rm SI}_N$ and $\sigma_{\rm ann}$ depend on several common parameters, 
including the Yukawa couplings $f_a$, $\tilde{f}_a$, and the masses $m_\psi$ and $m_{\eta_I}$.
In the numerical analysis, these parameters are varied within the ranges 
$0 < |f| < 1$, $-\pi < \arg(f_{a}), \arg(\tilde{f}_{a}) < \pi$, $5~{\rm [GeV]} < m_\psi < 100$[GeV], and $5~[{\rm MeV}] < m_{\eta_I} < m_\chi/2$ 
with $m_\chi$ fixed at $1.8~{\rm [GeV]}$.
Figure~\ref{fig:sigma} shows the resulting $\sigma^{\rm SI}_N$ as a function of $\lambda_{H\eta}$.
All shown points satisfy $\Omega_\chi < 0.01$ as required for a fully asymmetric DM scenario. 
Our results indicate that the quartic coupling is constrained to be $\lambda_{H\eta} \lesssim 10^{-3}$ 
to remain consistent with the fully ADM requirement.

\begin{figure}[htbp]
\centering
\includegraphics[width=0.45\textwidth]{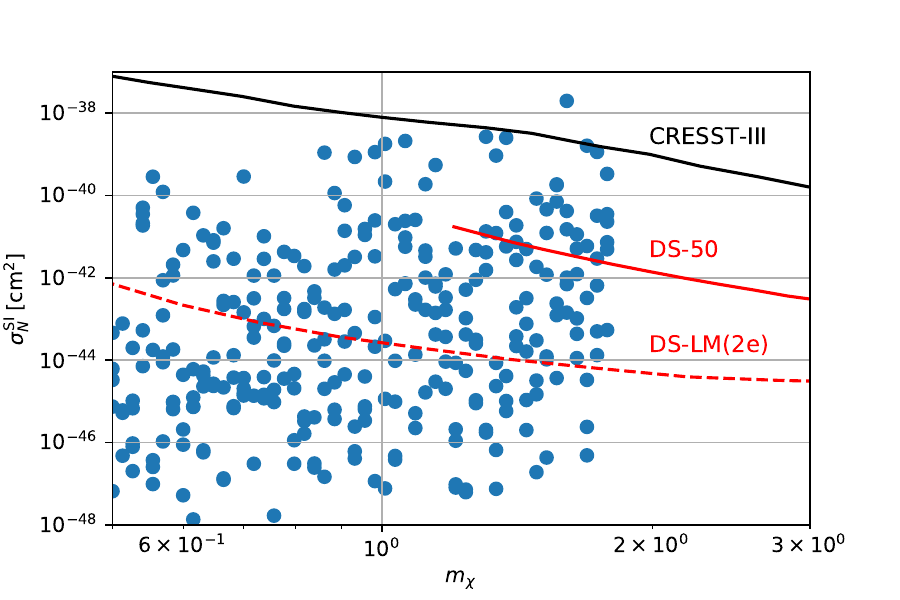} 
\caption{
Spin-independent nucleon--DM cross section for direct detection experiments 
as a function of the DM mass $m_\chi$ in the general case 
where both symmetric and asymmetric components contribute to the total dark matter abundance.
All points satisfy the observed relic density within the $3\sigma$ range.} 
\label{fig:sigma2}
\end{figure}

For the more general case, where both symmetric and asymmetric components contribute to the total dark matter abundance,
we study the correlation between $\sigma^{\rm SI}_N$ and the dark matter mass $m_\chi$ satisfying Eq.~\eqref{eq:mx}, 
i.e., reproducing the observed relic density $\Omega_{\rm DM}/\Omega_B \simeq 5$, with $\Omega_{\rm DM}=\Omega_{\rm asy} + \Omega_{\chi}$.
The results are shown in Fig.~\ref{fig:sigma2}, where we explore the same parameter space of $f_a$, $\tilde{f}_a$, $m_\psi$, and $m_{\eta_I}$.
All points shown satisfy the relic density constraint within the $3\sigma$ range,
$0.117 \leq \Omega_{\rm DM} h^2 \leq 0.123$.
To compare our predictions with experimental data, we overlay the current exclusion limits 
from the CRESST-III~\cite{CRESST:2019jnq} and DarkSide-50 (DS-50)~\cite{DarkSide-50:2022qzh} experiments.
Despite the loop-suppressed nature of the scattering process shown in Fig.~\ref{fig:SI},
a portion of the parameter space is already excluded by the DS-50 results.
On the other hand, for lighter dark matter masses, $m_\chi \lesssim 1~{\rm [GeV]}$,
current direct detection experiments lack sufficient sensitivity.
In this regime, the predicted scattering rates lie well above the so-called neutrino floor,
which represents a distinctive advantage of the ADM framework.
Future experiments such as DS-LM (sensitivity $\leq 2e$)~\cite{GlobalArgonDarkMatter:2022ppc},
designed to probe the sub-GeV dark matter region,
will be able to explore most of the presently allowed parameter space.
Hence, our model provides a testable target that could be either discovered or excluded by upcoming direct detection searches.

\begin{figure}[htbp]
\centering
\includegraphics[width=0.45\textwidth]{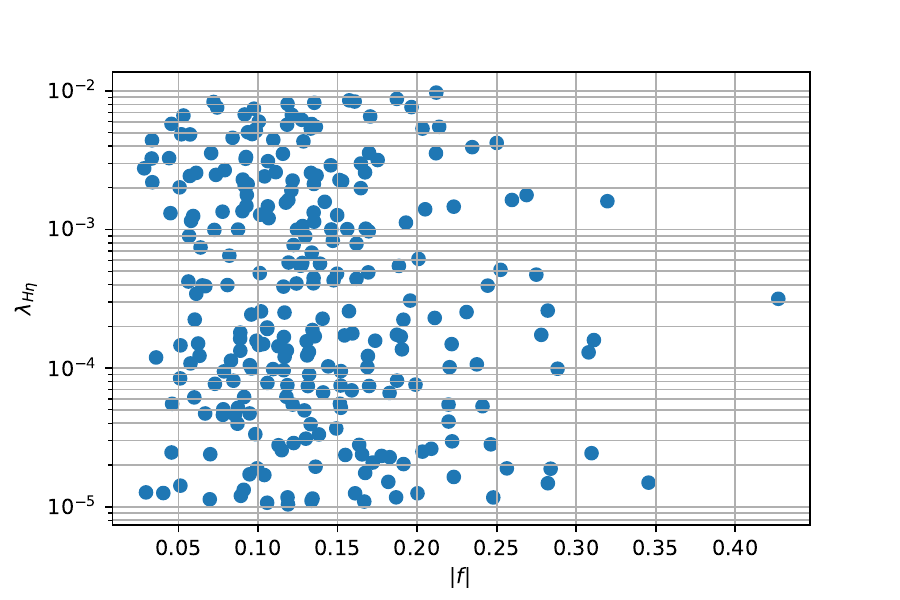} 
\caption{
Allowed parameter regions in the $(|f|, \lambda_{H\eta})$ plane, 
where all points satisfy both the DM relic density constraint within $3\sigma$ 
and the current limits from direct detection experiments shown in Fig.~\ref{fig:sigma2}.} 
\label{fig:lambda-f}
\end{figure}

Figure~\ref{fig:lambda-f} shows the allowed parameter regions in the $(\lambda_{H\eta}, |f|)$ plane.
All the plotted points satisfy both the observed dark matter relic abundance 
within the $3\sigma$ range and the current constraints from direct detection experiments.
As seen in the figure, a relatively broad region with $\lambda_{H\eta} \lesssim 0.01$ and $|f| < 0.1$ remains viable in the generic case.
This is because, for smaller $m_\chi$, the symmetric component 
can make a larger contribution to the total relic abundance, thereby relaxing the lower bound on $|f|$.

\subsection{Phenomenological Implications}

\subsubsection*{Higgs Invisible Decay Constraint}

Since we consider the case $m_ {\eta_I} \ll m_h/2$, 
the scalar potential term allows the Higgs boson to decay into a pair of $\eta_I$ scalars. 
This constrains the coupling constant $\lambda_{H\eta}$ in the interaction term 
$\lambda_{H\eta} H^\dagger H \eta^* \eta$ and the scalar mass $m_{\eta_I}$. 

The Higgs invisible decay width is given as
\begin{align}
\Gamma[h_0 \to \eta_I \eta_I] = \frac{\lambda_{H\eta}^2 v_{EW}^2}{32 \pi m_h} \sqrt{1 - \frac{4 m_{\eta_I}^2}{m_h^2}},
\end{align}
and the corresponding branching ratio is
\begin{align}
{\rm Br}[h_0 \to \eta_I \eta_I] = \frac{\Gamma[h_0 \to \eta_I \eta_I]}{\Gamma^{\rm SM}_{h_0} + \Gamma[h_0 \to \eta_I \eta_I]},
\end{align}
where $\Gamma^{\rm SM}_{h_0} = 4.6$ [MeV] is the SM Higgs total width~\cite{CMS:2024eka, CMS:2022ley}. 
The current upper limit on the Higgs invisible branching fraction is ${\rm Br}_{\rm invisible} \leq 14.5\%$ from ATLAS~\cite{ATLAS:2023dnm}, 
from which one can derive an upper bound $\lambda_{H\eta} \lesssim \mathcal{O}(10^{-2})$. 
This constraint is consistent with, and somewhat weaker than, the direct detection limit 
$\lambda_{H\eta} \lesssim 10^{-3}$ discussed in the previous section.

\subsubsection*{$\Delta N_{\mathrm{eff}}$}

Due to the Dirac nature of neutrinos, cosmological constraints on the effective number of relativistic degrees of freedom, $\Delta N_{\mathrm{eff}}$, must also be taken into account.  
The current experimental bound is $|\Delta N_{\mathrm{eff}}| < 0.17$ at the $1\sigma$ level~\cite{Planck:2018vyg}.  
In our framework, the SM Yukawa couplings to the right-handed neutrinos $\nu_R$ are highly suppressed.  
As a result, the interactions of $\nu_R$ with either SM particles or the dark sector are extremely weak, 
and the decoupling temperature of $\nu_R$ is much higher than the electroweak scale, as in conventional Dirac leptogenesis scenarios.  
The contribution of $\nu_R$ to the radiation density is suppressed by the entropy dilution factor,  
\begin{equation}
\Delta N_{\mathrm{eff}} = 3 \left( \frac{g_{*s}(T_{\nu_L})}{g_{*s}(T_{\nu_R})} \right)^{4/3},
\end{equation}
where $g_{*s}(T_{\mathrm{dec}})$ denotes the effective number of relativistic entropy degrees of freedom at the decoupling temperature $T_{\nu_{L/R}}$ of left- and right-handed neutrinos, respectively~\cite{Murayama:2002je}.  
Determining the precise decoupling temperature $T_{\nu_R}$ requires a dedicated analysis of the possible new interactions active near the heavy neutrino mass scale $v_S$. 
However, given the extremely small $\nu_R$ couplings in our model, it is reasonable to assume that $\nu_R$ decouples at or above $T \sim v_S$.
Under this assumption, the resulting contribution to the effective number of relativistic degrees of freedom is
\begin{equation}
\Delta N_{\mathrm{eff}}
= 
3 \left(
\frac{10.7}{
g_{*s}(\mathrm{SM} + \chi + \psi + \nu_R + N_{L,R} + S + \phi + \eta)}
\right)^{4/3}
\simeq 0.1,
\end{equation}
which is consistent with current cosmological observations.  
Future measurements, such as those anticipated from the CMB-S4 experiment~\cite{CMB-S4:2016ple}, aim to reach a sensitivity of $\Delta N_{\mathrm{eff}} \sim 0.03$.  
Such precision will allow this Dirac neutrino ADM scenario to be stringently tested, potentially validating or excluding it in the near future.

\section{Conclusion}\label{sec:conclusion}

In this work, we have proposed a new and simple framework for the Dirac neutrino scenario 
that naturally accommodates both the generation of the baryon asymmetry of the Universe and a viable asymmetric dark matter (ADM) candidate.  
Our model extends the Standard Model by introducing heavy Dirac neutrinos, a dark fermion $\chi$, heavier dark fermions $\psi_a$, and a complex scalar $\eta_I$,  
together with an extended $U(1)_X$ Froggatt-Nielsen-like mechanism and a discrete $\mathbb{Z}_4^D$ symmetry that stabilizes the dark matter.  
The $U(1)_X$ symmetry not only explains the hierarchical structure of the Yukawa couplings but also plays a central role in connecting the origins of neutrino and dark matter masses.
Importantly, the model achieves these goals with simple and anomaly-free charge assignments, without introducing any additional Standard Model charged scalars.  
As a result, it remains free from severe electroweak precision or flavor-changing constraints, ensuring a theoretically clean and predictive structure.

We have shown that the CP-violating decays of the lightest heavy Dirac neutrino can simultaneously generate asymmetries in both the Standard Model lepton sector and the dark sector.  
Through electroweak sphaleron processes, part of the lepton asymmetry is converted into the baryon asymmetry, 
successfully reproducing the observed baryon-to-photon ratio for natural choices of parameters.  
The dark matter relic abundance is directly tied to the baryon asymmetry, leading to a predictive ADM mass $m_\chi \simeq 2~\mathrm{[GeV]}$ consistent with cosmological observations.  
The symmetric component of dark matter efficiently annihilates into light scalars $\eta_I$ via Yukawa interactions, 
leaving only the asymmetric component and ensuring compatibility with Big Bang Nucleosynthesis constraints.  

The direct detection of $\chi$ proceeds through a one-loop Higgs-mediated spin-independent process, 
yielding a cross section that lies within the projected sensitivity of next-generation sub-GeV dark matter experiments.  
Furthermore, the hierarchical structure of couplings in our model may originate from higher-dimensional operators suppressed by a fundamental scale $\Lambda$, 
suggesting a possible Froggatt-Nielsen--like mechanism as the underlying source of small neutrino Yukawa couplings and dark sector mass parameters.  
Such an interpretation naturally connects the Dirac neutrino mass generation with flavor physics and points toward potential ultraviolet (UV) completions, 
for example within string-inspired frameworks featuring an additional $U(1)_X$ symmetry.
Overall, the proposed model provides a simple realization of Dirac leptogenesis with asymmetric dark matter, connecting the origin of neutrino masses, 
baryon asymmetry, and dark matter within a unified and testable framework.  
Future low-mass dark matter searches will be able to probe the predicted parameter space, 
offering a promising avenue to explore the deep connection between neutrino physics and the dark sector.

There are several interesting future directions.  
In our model, the sub-GeV dark matter (DM) scale arises naturally, offering a compelling framework that has recently 
attracted significant attention from both theoretical and cosmological perspectives.  
The DM candidate can exhibit self-interactions through a one-loop self-scattering process, $\chi \chi \to \chi \chi$, 
mediated by the heavier dark fermion $\psi$ and the scalar $\eta_I$.  
Such self-interactions may help to alleviate the so-called small-scale 
structure anomalies~\cite{Spergel:1999mh} (for a review, see~\cite{Tulin:2017ara} and references therein), which we plan to investigate in future work.  

It is also worthwhile to explore the broader implications of the new $U(1)_X$ gauge symmetry, as well as possible UV completions of the model.  
Given that the underlying physical scale $\Lambda$ in our setup is naturally expected to be close to the Planck scale, $M_{\rm{Pl}}$,  
it is reasonable to consider a string-theoretic origin of the $U(1)_X$ gauge theory.  
Radiative corrections in such a UV framework could induce higher-dimensional operators, 
potentially realizing a Froggatt-Nielsen-like mechanism that underlies the hierarchical structure employed in our construction.

\section*{Acknowledgements}
This work was supported by JST SPRING, Grant Number JPMJSP2115.
H.~O.~is supported in part by JSPS KAKENHI Grants No. 21K03554 and No. 22H00138.
S.~U.~is supported in part by JSPS KAKENHI Grant No. 22K14039.
The authors thank the Yukawa Institute for Theoretical Physics at Kyoto University.
Discussions during the YITP workshop YITP-W-25-10 on ``Progress in Particle Physics 2025"
were useful to complete this work.

\bibliographystyle{unsrt}

\bibliography{main}

\end{document}